\documentclass[aps,physrev,preprint,superscriptaddress]{revtex4-2}  % reprint vs preprint
\usepackage{graphicx}
\usepackage[bf]{caption} % For bolding the figure label 
\newcommand{\bcaption}[2]{\caption[#1]{\textbf{#1.} #2}} % Bolding first sentence of caption
\usepackage{amsfonts}
\usepackage{amsmath}
\usepackage{booktabs}

\begin{document}

\preprint{APS/Physics Review E.}

\title{\textbf{Temporally constraining source imaging estimates in an underdetermined neural system with eigenmodes of cortical geometry} 
}% 


% \author{Pok Him Siu}
% \email[]{pokhim.siu@student.unimelb.edu.au}
% %\homepage[]{Your web page}
% %\thanks{}
% %\altaffiliation{}
% \affiliation{%
% }%

\author{Pok Him Siu}
\email{pokhim.siu@student.unimelb.edu.au}
\affiliation{Department of Biomedical Engineering, The University of Melbourne, Australia}
\affiliation{Graeme Clark Institute, The University of Melbourne, Australia}
\author{Philippa J. Karoly}%
\affiliation{Department of Biomedical Engineering, The University of Melbourne, Australia}
\affiliation{Graeme Clark Institute, The University of Melbourne, Australia}
\author{Artemio Soto-Breceda}
\affiliation{Grenoble Insitut des Neurosciences, INSERM, France}
\author{Mark J. Cook}%
\affiliation{Department of Biomedical Engineering, The University of Melbourne, Australia}
\affiliation{Graeme Clark Institute, The University of Melbourne, Australia}
\affiliation{St Vincent's Hospital, Melbourne, Australia}
\author{David B. Grayden}
\affiliation{Department of Biomedical Engineering, The University of Melbourne, Australia}
\affiliation{Graeme Clark Institute, The University of Melbourne, Australia}


\date{\today}

\begin{abstract}
Geometric eigenmodes provide a compact and biologically grounded representation of large-scale neural activity. 
Previous work demonstrated that they can mitigate the underdetermined nature of electroencephalographic (EEG) and magnetoencephalographic (MEG) source localisation, an ill-posed inverse problem in which neural activity is reconstructed from non-invasive recordings. 
Beyond their spatial structure, neural field theory predicts the temporal evolution of eigenmodes through analytically derived transfer functions. 
Motivated by this framework, the present work investigates whether these transfer functions can be used to introduce temporal constraints into EEG source imaging. 
The approach is evaluated using simulated seizure dynamics generated by coupled Epileptor neural mass models.
% , as well as a coupled corticothalamic and hippocampal–septal neural field system. 
Transfer functions derived directly from neural field theory were found to be generally ineffective as temporal constraints for source localisation, primarily because they neglect cross-eigenmode coupling. 
Incorporating empirically estimated coupling terms substantially improves localisation performance, particularly in noisy conditions. 
Although estimating these eigenmode coupling interactions from experimental data remains challenging, the findings motivate dynamical source imaging approaches that combine spatial eigenmode structure with empirically informed cross-modal dynamics.

\end{abstract}

%\keywords{}

\maketitle


\section{Introduction \label{sec:intro}}

% Start with STATE OF THE ART NEURAL MODELLING (I would definitely combine neural mass models, and NFT into one paragraph. The next paragraph can then go into eigenmodes (with a clearer link)
The brain is a complex dynamical system in which large populations of interacting neurons collectively generate the neural activity observed experimentally. 
Population-level models reduce the immensly high dimensionality of neural dynamics through mean field and neural mass approximations, in which state variables, such as the mean firing rate or membrane potential, are tracked for neuronal populations rather than individual cells \cite{breakspear_dynamic_2017}. 
Neural field theory extends this framework into space by treating neural activity as a continuously varying field over the cortical surface \cite{deco_dynamic_2008, jirsa_field_1996}.
This enables a physiologically constrained description in which cortical activity propagates as damped waves and can be decomposed into natural spatial eigenmodes determined by the underlying spatial operator and cortical geometry \cite{gabay_cortical_2017, pang_geometric_2023}. 
Neural field theory, therefore, links the spatial structure of geometric eigenmodes with their temporal dynamics \cite{robinson_eigenmodes_2016, gabay_dynamics_2018}.
However, the cortical fields described by these models cannot be directly observed using non-invasive techniques such as electroencephalography (EEG) and magnetoencephalography (MEG).

% THIS IS A BETTER LEAD IN SENTENCE TO INTRODUCE THE PARAGRAPH TOPIC (SOURCE IMAGING) RIGHT AWAY
Source imaging seeks to infer this underlying cortical activity from EEG or MEG measurements. 
Source inference, an inverse problem, is underdetermined and ill-posed because the number of sensors, typically of order $10^2$, is much smaller than the number of candidate neural sources, typically of order $10^3$-$10^4$ \cite{he_electrophysiological_2018}. 
Hence, a source imaging algorithm requires both a forward model, which describes the mapping from cortical sources to sensors through volume conduction \cite{hallez_review_2007}, and additional constraints that select a stable solution to the inverse problem \cite{hauk_towards_2022}.
Although modern forward models can incorporate realistic head anatomy and tissue conductivities, the constraints used during inversion are often primarily motivated by mathematical or statistical considerations rather than by well-established models of neural dynamics \cite{he_electrophysiological_2018, kincses_modeling_1999}. 
Minimum-norm methods, for example, favour solutions with low overall source amplitude, while other approaches impose spatial smoothness \cite{pascual-marqui_functional_2002} or focality \cite{costa_sparse_2015}. 
Such assumptions can regularise the inverse problem effectively, but they do not explicitly describe how neural activity is organised and propagated across the cortex and can yield physiologically implausible source distributions \cite{haufe_large-scale_2011}. 
Neural field theory, therefore, offers a potential basis for constraining source reconstruction using the expected spatial and temporal structure of cortical dynamics.

% Previous work on application of biological priors / eigenmodes
Recent studies have demonstrated that structural eigenmodes of the human brain can provide effective spatial constraints for EEG source localisation and outperform many classic approaches \cite{siu_structural_2026, wang_geometry_2026, ruequeralt_connectome_2024}. 
Analogous to the decomposition of temporal signals into Fourier components, cortical activity can be represented as a weighted combination of structural eigenmodes. 
Each mode describes a spatial pattern constrained by either the brain's white matter connectivity \cite{ruequeralt_connectome_2024, siu_structural_2026} or cortical geometry \cite{wang_geometry_2026, siu_structural_2026}. 
Restricting source reconstruction to the subspace spanned by these modes reduces the dimensionality of the inverse problem while retaining physically and anatomically motivated spatial structure. 
Importantly, geometric eigenmodes achieve comparable localisation performance to connectome-derived modes while requiring substantially less anatomical information, making them an attractive spatial prior for source imaging \cite{siu_structural_2026}.

% Why focus on geometric eigenmodes and NFT.    
This work extends geometric eigenmode source imaging approaches by incorporating temporal dynamics prescribed by neural field theory, in addition to their intrinsic anatomical information, thereby imposing joint spatial and temporal constraints on the EEG inverse problem. 
Leveraging the distinct advantage that geometric eigenmodes arise naturally from the spatial operators used in neural field theory, we use established neural field formulations to describe their mode-dependent evolution and propagation analytically.
These predicted temporal dynamics provide an additional physiologically motivated constraint that is absent from purely spatial constraints in eigenmode source imaging. 

This paper is organised as follows. 
Sec.~\ref{sec:theory} introduces the theoretical background of neural field theory, geometric eigenmodes, and the derivation of the transfer function. 
Sec.~\ref{sec:algorithm} describes the algorithm for spatiotemporally constrained source inversion based on neural field theory. 
Sec.~\ref{sec:results} evaluates the performance of these temporally constrained eigenmodes on simulated seizures generated by a coupled Epileptor neural mass model, and a coupled cortico-thalamic and hippocampal-septal neural field model.
Finally, Sec.~\ref{sec:discussion} discusses the implications of the findings and directions for future work.

\begin{figure}
    \centering
    \includegraphics[width=\linewidth]{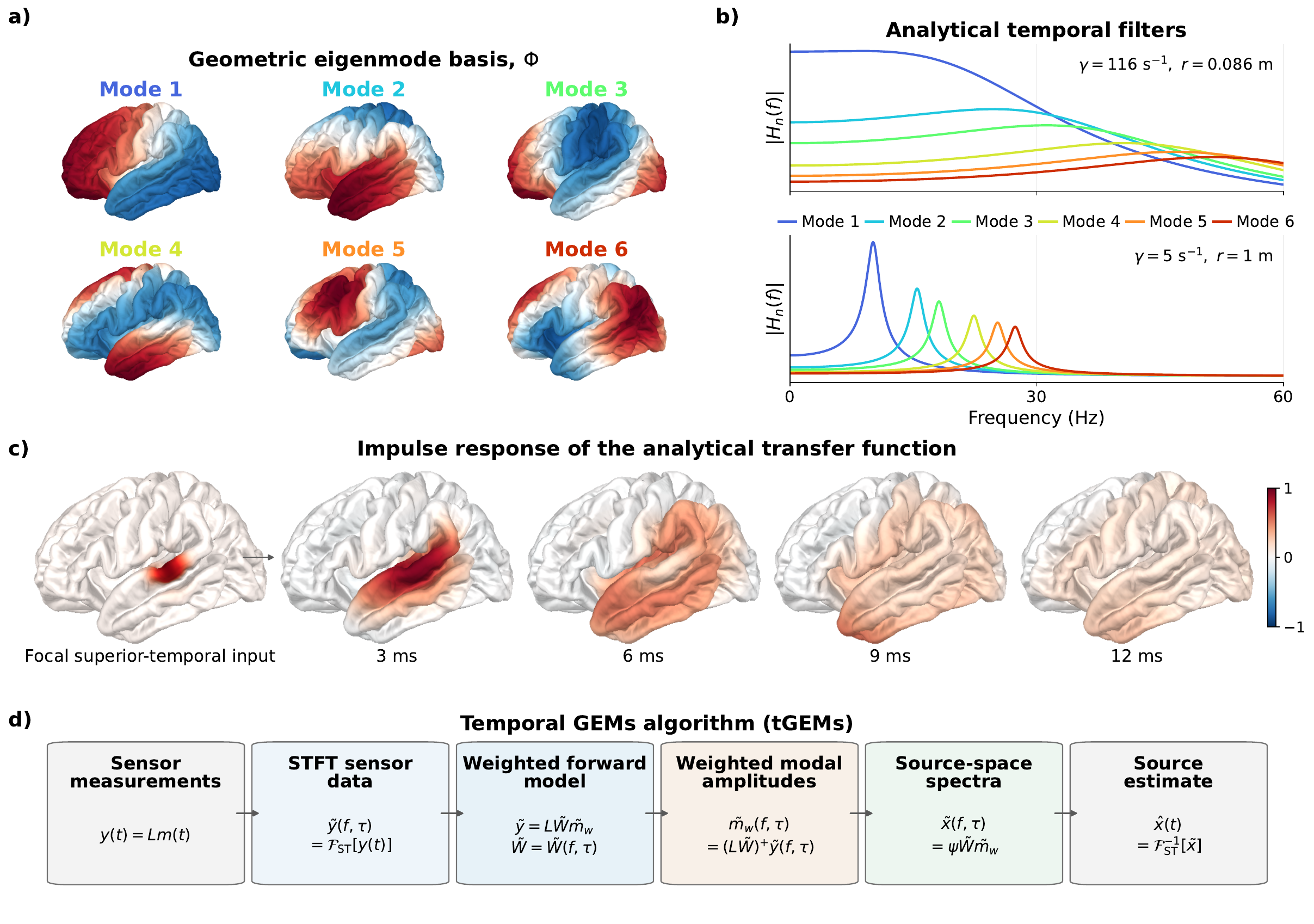}
    \bcaption{Temporal propagation of geometric eigenmodes}
    {a) The first 6 geometric eigenmodes, as computed through solving the eigenvalue problem on the surface of the left hemisphere.      
    b) The temporal response of the first 6 eigenmodes as derived analytically in Eq.~\ref{eq:transfer_fn_freq} with empirical parameters $\gamma=116$ and $r=0.86$ estimated from experiments \cite{gabay_cortical_2017, robinson_prediction_2001, gabay_dynamics_2018, phogat_unified_2025}, and select parameters $\gamma = 5$ and $r=1$, illustrating how responses change with different settings.  
    c) The causal response implied by the analytical transfer function for a focal superior-temporal input.
    d) Pipeline of the temporal GEM algorithm.
    }    
    \label{fig:tGEM_explanation}
\end{figure}

\section{Theory \label{sec:theory}}

\subsection{The Robinson Neural Field Model}

The Robinson neural field model provides a mesoscopic description of neural activity by modelling the propagation of mean firing activity through cortical tissue as a partial differential equation derived from average dendritic, somatic, and axonal physiology \cite{robinson_eigenmodes_2016}.
Originally developed as a physiologically grounded model of large-scale brain dynamics that retains analytical tractability \cite{robinson_propagation_1997}, it has since become a widely adopted framework for studying macroscopic brain dynamics, including normal brain rhythms and pathological activity such as epilepsy \cite{robinson_eigenmodes_2016, gabay_dynamics_2018, phogat_unified_2025}.
For the present work, this analytical tractability allows a closed-form transfer function in the spatial-frequency domain, enabling temporal dynamics to be naturally incorporated into the geometric eigenmode framework, as described in Section~\ref{sec:algorithm}.

In its simplest form, the model describes the spatiotemporal evolution of a scalar field, $\phi_{j}(\mathbf{x},t)$, typically interpreted as the mean axonal pulse density or synaptic activity \cite{robinson_propagation_1997} of the neural population ($\phi_{j}$) at position $\mathbf{x}$ and time $t$. 
Typically, different neural populations (denoted by the subscript $j$) are coupled together, such as an excitatory population ($\phi_{e}$), inhibitory population ($\phi_{i}$), a specific thalamic relay population ($\phi_{s}$), and a thalamic reticular population ($\phi_{r}$) \cite{gabay_dynamics_2018}.
The dynamics of the excitatory neural population ($\phi_{e}$) is the most important for the purposes of describing macroscopic cortical activity and is governed by an isotropic damped wave equation with finite propagation speed and no regenerative terms \cite{robinson_propagation_1997, robinson_eigenmodes_2016}, with
\begin{equation}
    \left( \frac{1}{\gamma_e^2} \frac{\partial^2}{\partial t^2} + \frac{2}{\gamma_e} \frac{\partial}{\partial t} + 1 - r_{e}^2 \nabla^2 \right) \phi_{e}(\mathbf{x},t) = Q_e(\mathbf{x},t) .
\label{eq:robinson_NFM}
\end{equation}
Here, $\gamma_e$ is a temporal damping rate that sets the characteristic decay time of activity proportional to the conduction velocity $v_e$ ($\gamma_e=v_e/r_e$), $r_{e}$ is a spatial length parameter representing the characteristic range of the excitatory axons, $\nabla^2$ is the Laplacian operator that captures the spatial coupling of the cortex, $Q_e(\mathbf{x},t)$ is the input to the system, comprising contributions from the wave, $\phi$, and other inputs external to the cortex (i.e., noise).  

For all other neural populations, the mean axonal ranges are sufficiently short so that $r$ is set to $0$, meaning that the activity converges to a local effect.  
The dynamics of these populations are spatially homogeneous and begin to resemble neural mass models, as described by
\begin{equation}
    \left( \frac{1}{\gamma_j} \frac{\partial}{\partial t} + 1 \right) ^2 \phi_{j}(\mathbf{x},t) = Q_j(\mathbf{x},t) .
\label{eq:robinson_NFM_other}
\end{equation}
This formulation allows for the generation of structured spatiotemporal patterns such as travelling waves, standing waves, and coherent oscillations across spatial scales \cite{robinson_propagation_1997, robinson_eigenmodes_2016}. 

\subsection{Geometric Eigenmodes}

Spatial eigenmodes ($\psi$) can be derived by solving the eigenvalue equation from the spatial Laplacian component of the excitatory neural field model operating on the surface mesh of the cortex \cite{gabay_cortical_2017, pang_geometric_2023},
\begin{equation}
    \nabla^2 \psi_k = -\lambda_k \psi_k .
\end{equation}
Here, $\nabla^2$ is the Laplace-Beltrami operator, which allows for a solution on the cortical surface, and $\psi$ represents the set of geometric eigenmodes with their corresponding eigenvalues $\lambda$.   
These eigenvalues and eigenmodes were ordered according to increasing spatial frequency such that $\psi_1$ was the eigenmode with the longest wavelength.  
Geometric eigenmodes were also grouped into distinct eigengroups of similar wavelengths of size $2\ell+1$, where $\ell$ is the $\ell$-th eigengroup.
In this work, the geometric eigenmodes were constructed on the
10 242 vertices white-matter surface mesh for each hemisphere (icosahedron 5 spacing).

\subsection{Analytical Transfer Function}

All neural activity (i.e., the synaptic activity $\phi(\textbf{x},t)$ and the input $Q(\textbf{x},t)$) can be expanded from the spatial eigenmodes with separated time components ($a_k(t)$ \& $q_k(t)$) \cite{gabay_cortical_2017, gabay_dynamics_2018},
\begin{equation}
    \phi(\mathbf{x},t) = \sum_k a_k(t) \psi_k(\mathbf{x}), \quad Q(\mathbf{x},t) = \sum_kq_k(t) \psi_k(\mathbf{x}).
    \label{eq:em_Expansion}
\end{equation}
Then, for a specific spatial eigenmode $\psi_k(\mathbf{x})$, its time-varying amplitude $a_k(t)$ can be given by
\begin{equation}
   \left( \frac{1}{\gamma^2} \frac{d^2}{dt^2} + \frac{2}{\gamma} \frac{d}{dt} + 1 + r_e^2 \lambda_k \right) a_k(t) = q_k(t).
   \label{eq:em_Independence}
\end{equation}
Under this linear, approximation, projection onto the orthogonal eigenbasis diagonalises the spatial operator, reducing the PDE to an independent ODE for each mode.

We can apply the temporal Fourier transform $\mathcal{F}$ and use the Fourier properties,
\begin{equation}
\mathcal{F}\left[ \frac{d}{dt} f(t) \right] = i\omega F(\omega), \quad
\mathcal{F}\left[ \frac{d^2}{dt^2} f(t) \right] = -\omega^2 F(\omega),
\end{equation}
to obtain
\begin{equation}   
\left( -\frac{\omega^2}{\gamma^2} + i \frac{2\omega}{\gamma} + 1 + r_e^2 \lambda_k \right)a_k(\omega) = q_k(\omega).
\end{equation}

Rearranging to obtain the relation $a_k(\omega) / q_k(\omega)$ then gives the analytical (${\mathrm{A}}$) transfer function for the $k$-th spatial eigenmode, 
\begin{equation}
    H_k^{\mathrm{A}}(\omega) = \frac{a_k(\omega)}{q_k(\omega)} = \frac{1}{-\omega^2/\gamma^2 + i 2\omega/\gamma + 1 + r_e^2 \lambda_k}.
\end{equation}

Finally, converting to Hz (as relevant to EEG/MEG) and letting $\omega = 2\pi f$ gives
\begin{equation} 
    \label{eq:transfer_fn_freq}
    H_k^{\mathrm{A}}(f) = \frac{\gamma^2 }{-4\pi^2f^2 + i \cdot 4\pi f\gamma + \gamma^2(1 + r_e^2 \lambda_k)},
\end{equation}
where the power spectral response is
\begin{equation}  
    |H_k^{\mathrm{A}}(f)|^2 = \left| \frac{\gamma^2 }{-4\pi^2f^2 + i \cdot 4\pi f\gamma + \gamma^2(1 + r_e^2 \lambda_k)}\right|^2.
\end{equation}

The power spectral response function $|H_k(f)|^2$ describes how strongly the $k$-th structural eigenmode amplifies or attenuates oscillatory inputs at frequency $f$.
Because each eigenmode has a different eigenvalue 
$\lambda_k$, the denominator of Eq.~\ref{eq:transfer_fn_freq} varies across modes, producing distinct spectral profiles.
Typically, modes associated with smaller eigenvalues correspond to large-scale spatial patterns and tend to produce stronger low temporal frequency responses, whereas modes with larger eigenvalues correspond to finer spatial structure and exhibit progressively attenuated responses.
These responses are shown in Fig.~\ref{fig:tGEM_explanation}b for eigenmode–eigenvalue pairs derived from Freesurfer's fsaverage subject \cite{fischl_freesurfer_2012}. 

\section{Algorithm \label{sec:algorithm}}
% This section uses array indexing notation of the form $M(f,t)$, where a multidimensional array $M$ (with dimension greater than two) is indexed by the frequency–time pair $(f,t)$ to produce the corresponding slice $M(f,t)$, which is at most a two-dimensional matrix. 
Throughout this section, $M(f,t)$ denotes the slice of array $M$ at frequency $f$ and time $t$, with the remaining dimensions forming the corresponding matrix.

\subsection{Geometric Eigenmode Source Localisation (GEM)}

For completeness, we summarise the static geometric eigenmode source inversion method introduced by Siu et al. (2026)~\cite{siu_structural_2026}. Under the quasi-static approximation, EEG measurements are related to cortical source activity through the linear forward model \cite{he_electrophysiological_2018, hallez_review_2007}
\begin{equation}
\label{eq:fwd}
y = Fx + \epsilon,
\end{equation}
where $F\in\mathbb{R}^{M\times N}$ is the lead-field matrix, $x\in\mathbb{R}^{N\times T}$ is the source activity, $y\in\mathbb{R}^{M\times T}$ is the measured EEG, and $\epsilon$ denotes measurement noise.
The dimensions represent the number of sensors ($M$), the number of possible sources in the brain ($N$), and the number of time steps ($T)$, while $\epsilon$ is a matrix that represents the overall measurement noise for each time step.

Representing the source activity in a basis of $K$ structural eigenmodes $\psi\in\mathbb{R}^{N\times K}$ gives
\begin{equation}
\label{eq:EEG_mode}
y = F\psi w + \epsilon,
\end{equation}
where $F\psi\in\mathbb{R}^{M\times K}$ contains the corresponding EEG modes and $w$ contains their coefficients. The least-squares estimate is
\begin{equation}
\label{eq:eeg_lstsq_sol}
\hat{w} = (F\psi)^+y,
\end{equation}
where $(\cdot)^+$ denotes the Moore–Penrose pseudoinverse.

Because low-order spatial modes are expected to dominate cortical activity \cite{pang_geometric_2023}, the eigenmodes are normalised and weighted using a diagonal matrix $W$ with entries
\begin{equation}
\label{eq:emode_diagbias}
W_i = \frac{1}{|\lambda_i|^\beta + c},
\end{equation}
where $\lambda_i$ is the eigenvalue associated with the $i$th mode. For geometric eigenmodes, we set $\beta=1/2$ and $c=2\pi$, yielding an approximately inverse-wavenumber weighting that favours spatial scales more readily observable with EEG. 

Therefore, the reconstructed source activity is
\begin{equation}
\label{eq:source_lstsq_sol}
\hat{x} = \psi W(F\psi W)^+y.
\end{equation}

To limit the amplification of measurement noise by small singular values \cite{sekihara_electromagnetic_2015}, the pseudoinverse is computed using truncated singular value decomposition. 
Unless otherwise stated, the relative truncation threshold is set to $r_\text{cond} = \mathrm{min}(10^{-\mathrm{SNR}/10},10^{-1.2})$, consistent with the regularisation scale commonly used in minimum-norm source reconstruction. 
The choice of minimum is to avoid the truncation of too many modes \cite{siu_structural_2026}. 

\subsection{The Temporal GEM Algorithm (tGEM) \label{sec:tempgem_eqs}}

To incorporate frequency information into inverse source imaging problem through geometric eigenmodes, we start with a modified version of Eq.~\ref{eq:EEG_mode},
\begin{equation} \label{eq:time_sensors}
    y(t) = Lm(t),
\end{equation}
where, as before, $y(t)$ are the sensor measurements and $m(t)$ are the amplitudes of the eigenmodes for each time point.
For simplicity of notation, we have dropped the sensor error $\epsilon$  and made the substitution $L = F\psi$.

Using the short-time Fourier transform $\mathcal{F}_{ST}$ takes us to a windowed time $\tau$ and frequency $f$ space,
\begin{equation}
    \label{eq:stft_gemfwd}
    \tilde{y} (f,\tau) = \mathcal{F}_{ST}\left[Lm(t) \right] = L \tilde{m}(f,\tau),
\end{equation}
where $\tilde{y}\in\mathbb{C}^{M\times F \times T}$ and $\tilde{m}\in\mathbb{C}^{K\times F \times T}$ are the Fourier-transformed versions of their respective variables and functions of STFT time windows and frequency.  
An implicit window size ($64$ time points) and hop length ($32$ time points) are assumed for the short-time Fourier transform.
We denote $\tilde{y}(f,\tau)\in\mathbb{C}^{M}$ and $\tilde{m}(f,\tau)\in\mathbb{C}^{K }$ as vectors at a specific time window $\tau$ and frequency $f$ to remain within matrix notation. 

As in Eq.~\ref{eq:source_lstsq_sol}, we can apply a weighting matrix $\tilde{W}\in\mathbb{C}^{K\times K\times F \times T}$ to the mode amplitudes,
\begin{equation}
    \label{eq:tgem_weighted}
    \tilde{y} (f,\tau) = L \tilde{W}(f,\tau) \tilde{m}_w(f,\tau) ,
\end{equation}
where we define the transfer function weighting matrix $\tilde{W}(f,\tau)$ to be stationary within each STFT window but may vary between windows and $\tilde{m}_{w}(f,\tau)$ are the coefficients in the weighted modal basis, such that  $\tilde{m}(f,\tau)=\tilde{W}(f,\tau) \tilde{m}_w(f,\tau)$. 
Variations of the weighting matrix constructed from the cortical transfer functions are described in Sec.~\ref{sec:algo_weighting_matrix}. 

The weighted modal amplitudes, $\tilde{m}(f,\tau)_w$, can then be solved in Fourier space by
\begin{equation}
    \tilde{m}_w(f,\tau) = (L\tilde{W})^{+}\tilde{y}(f,\tau) ,
\end{equation}
where, for a piecewise-stationary transfer matrix, the pseudoinverse need only be recomputed when 
$\tilde {W}(f,\tau)$ changes.

Converting the weighted modal amplitudes to source-space Fourier amplitudes ($\tilde{x}$) for each time point gives 
\begin{equation}
    \tilde{x} = \psi\tilde{W}\tilde{m}_w = \psi\tilde{W}(L\tilde{W})^{+}\tilde{y}.
\end{equation}

Taking the inverse short-time Fourier transform $ \mathcal{F}^{-1}_{ST}$ then provides a solution to the source inversion problem,
\begin{equation}
    \hat{x} = \mathcal{F}^{-1}_{ST} \left[\psi\tilde{W}(L\tilde{W})^{+}\tilde{y} \right].
    \label{eq:tgems}
\end{equation}
From this point forward, we refer to the geometric eigenmode source inversion technique (described by Eq.~\ref{eq:source_lstsq_sol}) as GEM.  
We refer to the transfer function-weighted geometric eigenmode source inversion technique, which incorporates temporal (frequency) information (described by Eq.~\ref{eq:tgems}), as tGEM.    

\subsection{transfer function Weighting Schemes
\label{sec:algo_weighting_matrix}}

We consider several constructions of the transfer function weighting matrix, $\tilde{W}(f,\tau)\in\mathbb{C}^{K\times K}$. 
Each construction specifies how latent coefficients in the weighted modal basis contribute to the cortical modal amplitudes at frequency $f$ and STFT frame $\tau$.

\subsubsection{Analytical Transfer Function }

The analytical transfer function in Eq.~\ref{eq:transfer_fn_freq} assigns a complex frequency response $H_k^{\mathrm{A}}(f)$ to the eigenmode-eigenvalue pair $(\psi_k,\lambda_k)$. 
Because the analytical model treats the eigenmodes as independent, it produces a diagonal weighting matrix,
\begin{equation}
\label{eq:diag_weight_tf}
    \tilde{W}^{\mathrm{A}}(f) =
    \begin{bmatrix}
        H_1^{\mathrm{A}}(f) & 0 & \cdots & 0 \\
        0 & H_2^{\mathrm{A}}(f) & \cdots & 0 \\
        \vdots & \vdots & \ddots & \vdots \\
        0 & 0 & \cdots & H_K^{\mathrm{A}}(f)
    \end{bmatrix},
\end{equation}
which depends on frequency but remains constant across STFT frames.

\subsubsection{Empirical Transfer Function}

The analytical construction assumes that the linear cortical operator remains diagonal in the eigenmode basis. 
However, nonlinear dynamics and cortico-subcortical feedback can introduce statistical dependencies between modes. 
Therefore, we construct an empirical weighting matrix that contains non-zero off-diagonal entries.

To construct this empirical weighting matrix, we first project the known reference source activity $x^{\mathrm{ref}}(t)$ onto the eigenmode basis to obtain the empirical weights of each mode,
\begin{equation}
    m^{\mathrm{ref}}(t)
    =
    \psi^{+}x^{\mathrm{ref}}(t).
\end{equation}
The STFT is applied to obtain $\tilde{m}^{\mathrm{ref}}_k(f,\tau)$, the STFT coefficient of mode $k$.  
For each ordered mode pair $(j,i)$, we estimate the cross-spectrum,
\begin{equation}
\label{eq:empirical_cross_spectrum}
    S_{ji}(f)
    =
    \left\langle
        \tilde{m}^{\mathrm{ref}}_j(f,\tau)
        \tilde{m}^{\mathrm{ref}*}_i(f,\tau)
    \right\rangle_{\tau},
\end{equation}
where $(\cdot)^*$ denotes complex conjugation and $\langle\cdot\rangle_{\tau}$ denotes an average across STFT frames. 
We then define the empirical transfer weight from reference mode $i$ to output mode $j$ as
\begin{equation}
\label{eq:empirical_transfer_weight}
    H^{\mathrm{E}}_{ji}(f)
    =
    \frac{S_{ji}(f)}
         {S_{ii}(f)+\varepsilon},
\end{equation}
where $\varepsilon$ stabilises the estimate when the reference-mode power is small. 
Repeating this calculation with each mode as the reference mode gives the full empirical weighting matrix,
\begin{equation}
\label{eq:empirical_weight_matrix}
    \tilde{W}^{\mathrm{E}}(f)
    =
    \begin{bmatrix}
        1 & H^{\mathrm{E}}_{12}(f) & \cdots & H^{\mathrm{E}}_{1K}(f) \\
        H^{\mathrm{E}}_{21}(f) & 1 & \cdots & H^{\mathrm{E}}_{2K}(f) \\ 
        \vdots & \vdots & \ddots & \vdots \\
        H^{\mathrm{E}}_{K1}(f) & H^{\mathrm{E}}_{K2}(f) & \cdots &1
    \end{bmatrix}.
\end{equation}
Under this formulation, the diagonal entries, which describe normalised within-mode responses, are approximately unitary.

\subsubsection{Hybrid Transfer Function}
The hybrid transfer function combines the analytical within-mode frequency responses with the cross-mode structure of the empirical matrix.
To achieve this, each empirical column is scaled by the corresponding analytical transfer function such that each element of the matrix is of form
\begin{equation}
    \label{eq:hybrid_column_scaled}
    \tilde W_{ji}^{\mathrm{HC}}({f,\tau}) = H^E_{ji}(f,\tau) H^A_i(f).
\end{equation}

Further variants of the empirical transfer function matrix were also evaluated.
\paragraph{Specific}
For each Epileptor simulation, an empirical transfer function matrix was estimated by averaging across its STFT frames. 
The resulting simulation-specific matrix was then used to reconstruct that same simulation.

\paragraph{Averaged}
A single empirical transfer function matrix was obtained by averaging the simulation-specific matrices across all Epileptor simulations. 
This common matrix was then applied to every simulation.

\paragraph{Oracle}
To retain the temporal variation of the empirical transfer function, the STFT frames in Eq.~\ref{eq:empirical_cross_spectrum} were not averaged. 
This yielded a time-dependent matrix $\tilde{W}^{\mathrm{E}}(f,\tau)$, where $\tau$ indexes the STFT frame. 
The reconstruction was then performed using the corresponding frame-specific transfer function, with $H^{\mathrm{E}}_{ji}(f)$ replaced by $H^{\mathrm{E}}_{ji}(f,\tau)$.
Because this approach uses the transfer function associated with each time frame of the simulated source activity, it represents an oracle case rather than a practically available reconstruction method \cite{vincent_oracle_2007}.

\section{Results \label{sec:results}}

To evaluate the source localisation algorithms, coupled neural mass Epileptor simulations were used to generate a ground truth and a corresponding EEG data set, as described by Siu et al. (2026)~\cite{siu_neural_2026}.
The coupled Epileptor models represent the underlying sources of neural activity.
To construct the corresponding simulated EEG data, a forward matrix $F$ is required to transform the neural sources.
For this study, $F$ was constructed with a head volume conduction model computed using a three-shell Boundary Element Method (BEM) \cite{he_electric_1987} approach implemented in \textsc{OpenMEEG} and \textsc{MNE-Python} \cite{gramfort_openmeeg_2010, gramfort_mne_2014}.
$M$ refers to the number of sensors, which was 88 electrodes for the 10-10 spacing montage used for this study.
$N$ is the number of sources, which was set to 20,484 dipoles across the entire cortical surface, representing an average spacing close to 3.1~mm between dipole sources.
In total, the performances of the different geometric eigenmode source localisation algorithms were evaluated across $n=2038$ Epileptor simulations of different seizure conditions.
Further model and simulation details are provided in App.~\ref{app:sim_epileptor}.

\begin{figure} 
    \centering
    \includegraphics[width=\linewidth]{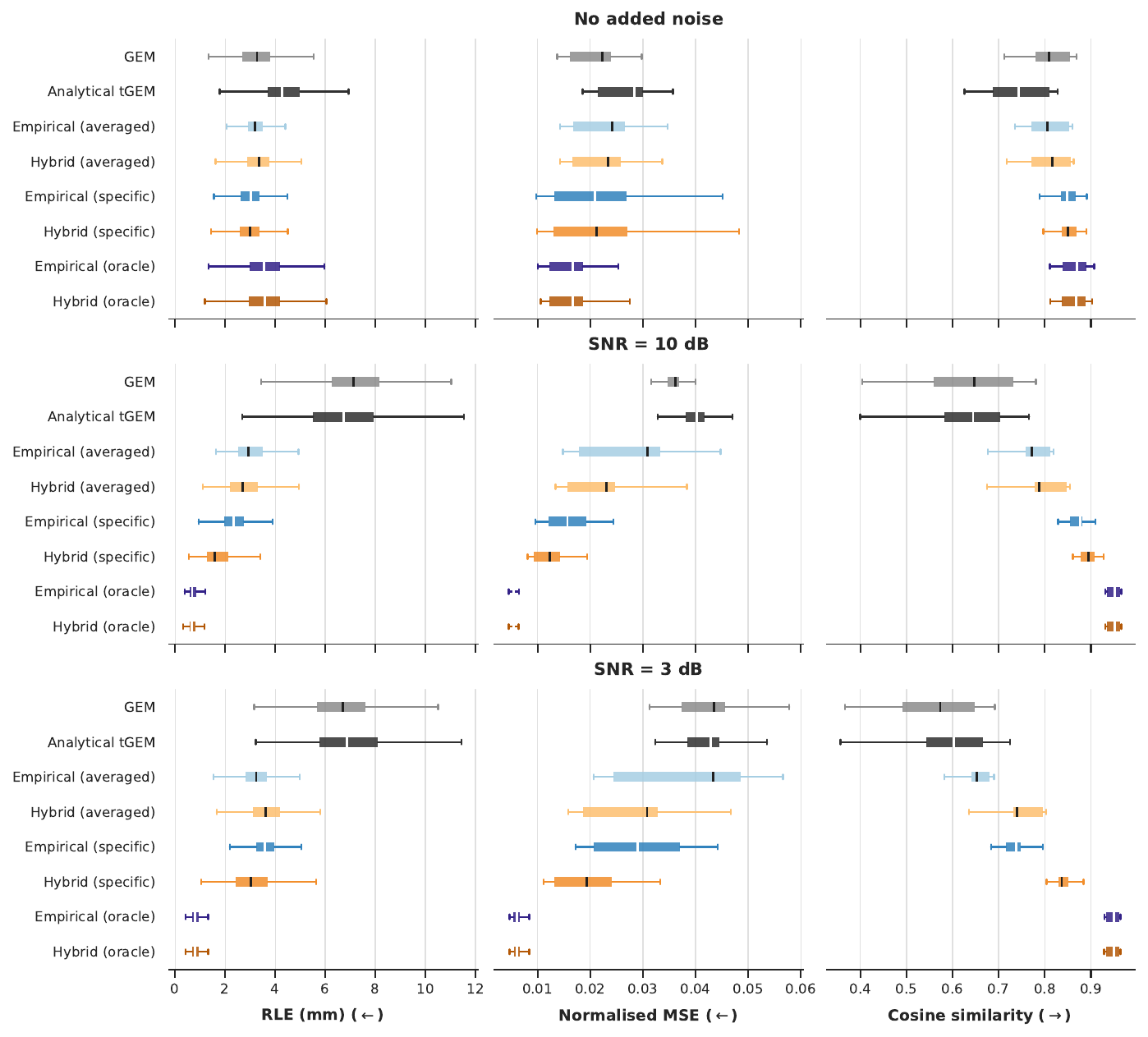} 
    \bcaption{Performance of geometric eigenmode source localisation approaches}
    {Boxplots signifying the range, interquartile range and median of the algorithms across all $n=2038$ Epileptor simulations.
    The colour of the boxplot corresponds to a different tested variation of the geometric eigenmode source localisation approach.
    Calculated signifies that the empirical transfer function was calculated from the corresponding simulation, whereas average signifies that the empirical transfer function used was the averaged across all simulations.
    Each column depicts a different metric: RLE, Normalised MSE, Cosine similarity, where the arrow signifies whether better performance is a lower number (left arrow), or a higher number (right arrow). 
    Each rowed subplot is a different signal to noise ratio: No added noise, 10 dB, 3 dB.     
    }
    \label{fig:tgem_main_analysis}
\end{figure}

Fig.~\ref{fig:tgem_main_analysis} shows the time-averaged reconstruction metrics over the full seizure.
To assess whether empirically informed transfer functions could provide useful temporal constraints and how their utility depends on the specificity of the available dynamical information, we considered three progressively more informative estimates: an averaged transfer function shared across simulations, a simulation-specific transfer function fixed over time, and a time-varying oracle transfer function derived from the underlying source dynamics. 
Hybrid variants combining the empirical cross-modal relationships with the mode-dependent scaling of the analytical transfer function were also evaluated.
Reconstruction metrics capture different performance domains (spatial, temporal, general similarity) via the region localisation error (RLE), mean-squared error (MSE), and cosine similarity (for detailed derivations of each metric, refer to App.~\ref{app:metrics}). 
The corresponding paired comparisons with GEM as the reference are reported in Tab.~S1. %~\ref{tab:gem_paired_comparisons}
Because the large number of simulations can render even small paired differences statistically significant, we used Cohen's $d_z$ when comparing each method with GEM.  
Cohen's $d_z$ is calculated as the mean paired difference divided by the standard deviation of the paired differences.

Although the main results use Epileptor simulations, the overall findings were reproduced in a set of three coupled cortico-thalamic and hippocampal-septal neural field model simulations, which were fitted to real patient seizures, available as Sup.~S2.%\ref{sec:nfm_results}.

\subsection{Performance of Temporal Geometric Eigenmode Source Localisation (tGEM)
\label{sec:results_tgem}}

The analytical tGEM did not provide a consistent improvement over GEM; 
in the absence of added noise, it performed substantially worse across all three metrics, with large negative effects for RLE ($d_z=-0.947$), normalised MSE ($d_z=-4.566$), and cosine similarity ($d_z=-2.502$). 
At $\mathrm{SNR}=10~\mathrm{dB}$, analytical tGEM produced a small improvement in RLE ($d_z=0.294$), but remained worse for MSE ($d_z=-1.820$) and cosine similarity ($d_z=-0.298$). 
At $\mathrm{SNR}=3~\mathrm{dB}$, the effect on MSE was negligible ($d_z=-0.107$) and RLE was slightly worse ($d_z=-0.222$), although cosine similarity improved substantially ($d_z=1.548$).
Thus, applying a fixed analytical transfer function did not reliably improve reconstruction across metrics and noise levels.

Given the limited performance of the analytical tGEM, we next asked whether empirically estimated transfer functions could improve reconstruction.
Without added noise, the averaged empirical and hybrid approaches provided negligible changes in RLE and reduced MSE performance relative to GEM, although the hybrid formulation was less detrimental than the purely empirical formulation ($d_z=-0.667$ compared with $-0.824$). 
With added noise, however, both averaged approaches improved all three metrics. 
At $\mathrm{SNR}=10~\mathrm{dB}$, these improvements were large for both methods, with the hybrid approach producing stronger effects for MSE and cosine similarity, whereas the empirical approach produced a slightly stronger improvement in RLE. 
At $\mathrm{SNR}=3~\mathrm{dB}$, the same pattern was observed, with empirical averaging yielding large improvements in RLE and cosine similarity but only a medium improvement in MSE, while hybrid averaging produced large improvements across all metrics. 
Estimating the specific transfer functions separately for each simulation further improved performance. 
Under noise-free conditions, the simulation-specific empirical and hybrid approaches produced small improvements in RLE, negligible changes in MSE, and large improvements in cosine similarity. 
At both noise levels, all effects were large, with the simulation-specific hybrid method giving the strongest non-oracle improvements in MSE and cosine similarity while performing similarly to the empirical method for RLE.

The oracle methods provided the best overall reconstruction performance. 
Without added noise, both oracle formulations produced large improvements in MSE and cosine similarity, although RLE remained effectively unchanged relative to GEM. 
At $\mathrm{SNR}=10$ and $3~\mathrm{dB}$, the oracle methods substantially outperformed GEM across all metrics, with effect sizes ranging from large ($d_z=2.819$) to very large ($d_z=8.683$). 
The empirical and hybrid oracle results were nearly indistinguishable, indicating that both formulations converge when the correct mode-dependent transfer functions are known. 
The apparent improvement in oracle performance after the addition of noise, particularly for RLE and MSE, is due to the dependence on the SNR for selecting the number of modes truncated during the pseudoinverse.
This is discussed further in Sec.~\ref{sec:discussion}. 

\begin{figure}[htbp]
    \centering
    \includegraphics[width=0.9\linewidth]{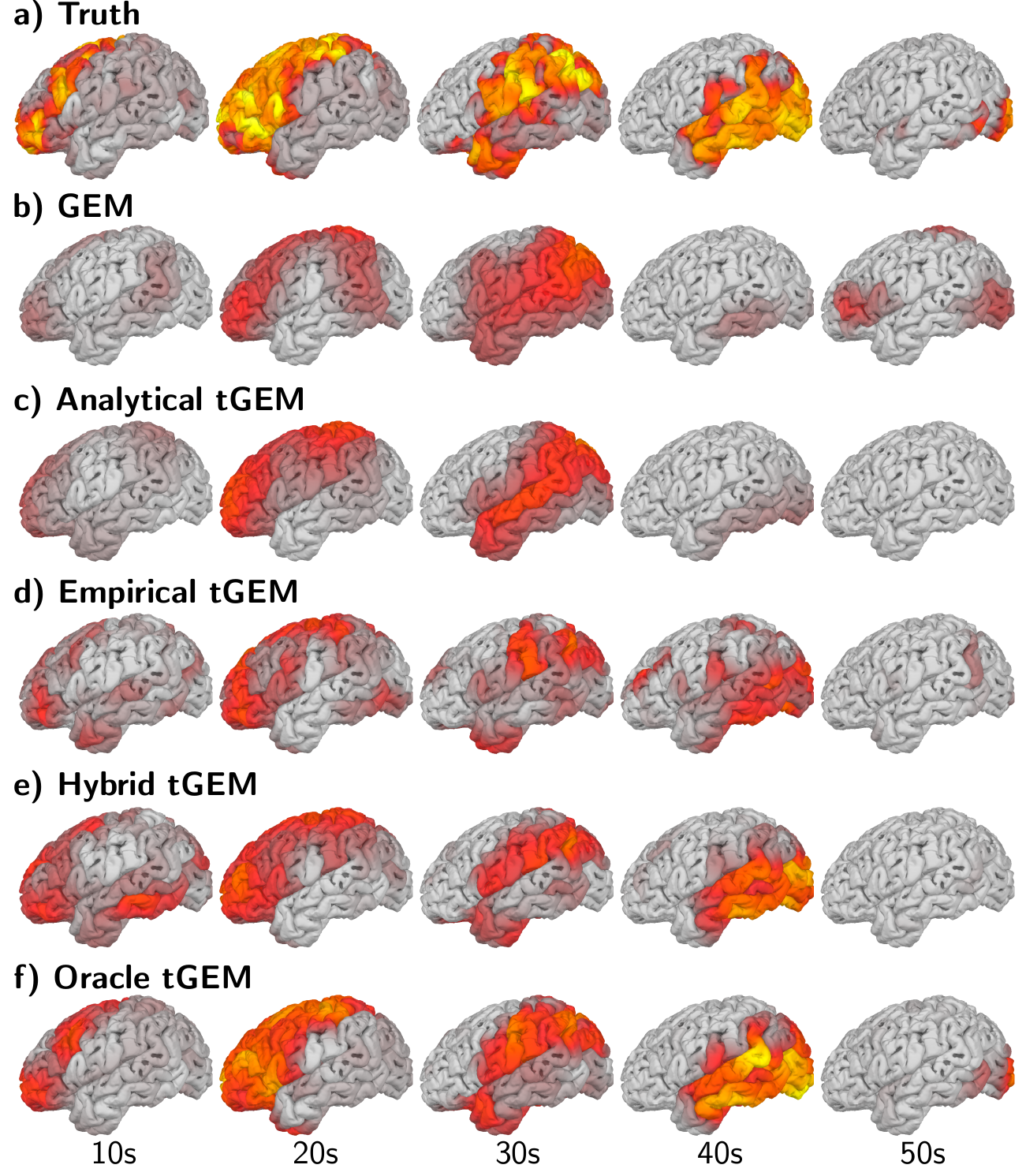}
    \bcaption{Source estimates of thalamic seizure with high noise (3~dB SNR) }
    {Cortical activity is shown at t=10, 20, 30, 40, and 50 s for a) the simulated ground truth, b) GEM, c) analytical tGEM, d) empirical tGEM, e) hybrid tGEM, and f) oracle tGEM. 
    Brighter colours (yellow) indicate greater source activity, while red regions indicate moderate activity, and grey (or transparent) regions indicate no activity. 
    The empirical and hybrid reconstructions use simulation-specific transfer functions. 
    All reconstructions were obtained from the same representative Epileptor simulation of a seizure originating from the thalamus with connectivity strength of $g = 2$ after $3~\mathrm{dB}$ of Gaussian white noise was added to the signal.}
    \label{fig:stc_estimates_3db}
\end{figure}

To analyse the performance of each algorithm across time, the estimates for a representative thalamic seizure with 3~dB noise (high noise) are presented in Fig.~\ref{fig:stc_estimates_3db}. 
Under this high-noise condition, GEM and analytical tGEM recover the spatial pattern reasonably well when the activity is widespread, but perform poorly when the true activity is more spatially restricted, particularly at 10, 40, and 50~s.
The stationary empirical approaches (labelled empirical and hybrid tGEM) recover more of the underlying activity at 10 and 40~s, although their performance also deteriorates at 50~s, when the signal is weakest relative to the noise.
In contrast, the oracle tGEM most consistently reproduces the location and spatial extent of the simulated activity across time.
The corresponding source estimates without added noise are shown in Fig.~S1. % ~\ref{fig:stc_estimates_nonoise}. 
Removing measurement noise improves all methods and reduces the differences between them, although the empirical, hybrid, and oracle approaches generally remain closer to the simulated activity than GEM and analytical tGEM.
Overall, these results show that the benefit of temporal constraints depends on both the accuracy of the transfer function estimate and the measurement noise. 
Oracle transfer functions provide the strongest reconstruction performance, while simulation-specific hybrid transfer functions offer the most effective non-oracle alternative.

\subsection{Computational cost and memory requirements}

\begin{figure}[htbp]
    \centering
    \includegraphics[width=0.5\linewidth]{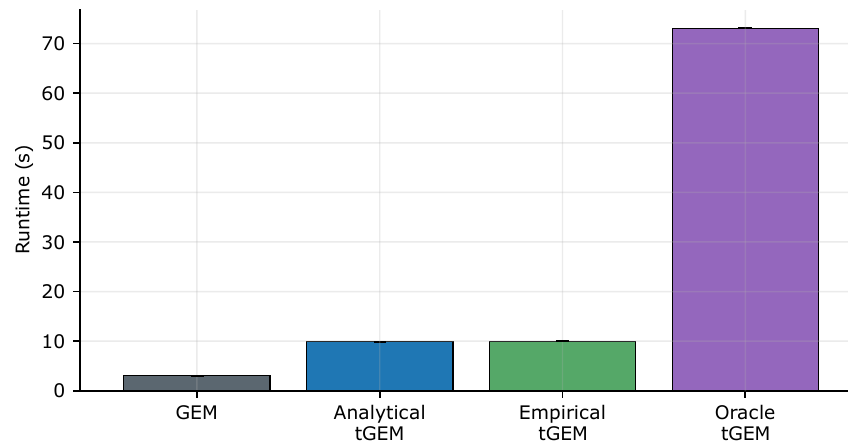}
    \bcaption{Computational runtime of the source reconstruction algorithms}{
    Runtime required by GEM (grey), analytical tGEM (blue), empirical tGEM (green), and oracle tGEM (purple) to reconstruct a full 60~s seizure using a cortical surface source space comprising 20,484 dipoles. Benchmarks were performed on a MacBook Pro with an Apple M1 Pro processor.
    }
    \label{fig:benchmark_runtimes}
\end{figure}

Fig.~\ref{fig:benchmark_runtimes} compares the computational runtime of each algorithm to reconstruct a full 60~s seizure. 
For the cortical surface source space comprising 20,484 dipoles, GEM required approximately $0.4~\mathrm{ms}$ per time step, while the current tGEM implementation required approximately $2.1~\mathrm{ms}$ per time step. 
By contrast, the oracle implementation required longer than the duration of the input recording to complete the reconstruction and, therefore, is not currently suitable for real-time operation without specialised hardware or further optimisation.
The runtimes of the non-oracle methods indicate that real-time processing may be feasible for some macroscopic EEG applications depending on the sampling rate, permitted latency, and computational demands of the remaining processing pipeline. 
Implementing a nonstationary transfer function in real time would require further code optimisation or less frequent updates of the transfer function estimate, rather than recalculating it at every time point.
Although offline processing is sufficient for retrospective applications, such as epileptogenic-zone localisation, real-time source estimation would support applications including long-term seizure monitoring and brain-computer interfaces.
The memory requirements of tGEM are primarily determined by the short-time Fourier-transform representation and, therefore, scale with the number of retained frequency bins, time windows, and eigenmodes.
As such, the choice of STFT parameters introduces a trade-off between the temporal resolution, frequency resolution, and computational resource requirements.

\section{Discussion \label{sec:discussion}}
The key finding of this study is that temporal information can improve geometric eigenmode source localisation, but the interactions between modes must be represented accurately.
Empirically estimated transfer functions substantially improved reconstruction, particularly when cross-eigenmode interactions were retained and the estimates were tailored to individual simulations or time windows. 
The oracle formulation provided the strongest performance, representing the upper limit of the potential benefit of accurately capturing time-varying modal relationships. 
In contrast, the analytical neural field transfer function did not consistently improve localisation when applied as an independent temporal constraint. 
This suggests that a fixed description, in which spatial eigenmodes evolve independently, is too restrictive across the dynamical regimes considered here.
The following discussion details the specific mathematical and physiological assumptions of the framework, their implications for the observed results, and suggestions for future work on how empirical transfer functions could be estimated in practice. 

\subsection{Limitations and Assumptions of the Analytical Framework}
\subsubsection{Linear neural field theory}
The analytical framework is fundamentally predicated on the assumptions of linear neural field theory, specifically the use of orthogonal spatial eigenmodes and linear transfer functions. 
These mathematical constructs assume that the underlying neural system operates in a regime where wave-like behaviour occurs and that local linearisations are valid. 
However, seizures and other brain activity are inherently nonlinear and non-stationary phenomena. 
Since the assumptions are necessarily violated in longer-term data, the utility of the framework is confined to specific windows of applicability.
To accommodate temporal nonstationarity, the formulation in Sec.~\ref{sec:tempgem_eqs}, particularly Eq.~\ref{eq:stft_gemfwd}, uses the short-time Fourier transform (STFT) to approximate the signal as locally stationary within finite time windows. 
The choice of STFT parameters represents an additional modelling assumption. 
In particular, window and hop sizes determine the trade-off between frequency and temporal resolution, respectively; a broader exploration of these parameters may further improve reconstruction performance.

\subsubsection{Stationarity of the transfer function}
The frequency-domain formulation assumes approximate stationarity within each STFT window, allowing the dynamics to be represented using transfer functions.
A stronger assumption is introduced in the analytical and non-oracle tGEM approaches, where the same transfer function weighting is applied throughout the seizure (Eq.~\ref{eq:tgem_weighted}). 
Thus, although the source signal is treated as locally stationary, the relationship between spatial eigenmodes and their temporal dynamics is itself assumed to remain stationary over considerably longer time scales.
This assumption is unlikely to hold universally during seizures, which can involve rapid transitions between dynamical regimes and substantial changes in spectral content. 
In the analytical transfer function of Eq.~\ref{eq:transfer_fn_freq}, the temporal response is governed primarily by the propagation parameters $\gamma$ and $r$, for which approximate empirical values have been previously reported \cite{gabay_cortical_2017,robinson_prediction_2001,gabay_dynamics_2018,phogat_unified_2025}. 
However, fixed values of these parameters cannot capture temporal variation in the effective dynamics of the underlying system. 
This limitation is consistent with the improved reconstruction obtained using the time-varying oracle transfer functions.

\subsubsection{Independent evolution of eigenmodes}
An additional assumption arises from the independent treatment of the spatial eigenmodes. 
In deriving the analytical transfer functions, the input is decomposed into cortical spatial eigenmodes  (Eq.~\ref{eq:em_Expansion}), and the linearised dynamics allow each mode to evolve independently according to its own transfer function (Eq.~\ref{eq:em_Independence}). 
This produces the diagonal transfer function representation used in Eq.~\ref{eq:diag_weight_tf}, in which activity within one eigenmode does not directly contribute to the temporal response of another.
This assumption becomes increasingly restrictive when recurrent coupling, heterogeneous inputs, or nonlinear population dynamics are present. 
Changes in excitatory gain, thalamocortical feedback, or cortico-hippocampal coupling can alter the effective response of the system and introduce interactions between spatial modes \cite{phogat_unified_2025}. 
As the system approaches more strongly nonlinear dynamical regimes, recurrent interactions can thus redistribute activity across eigenmodes, breaking the assumption of independent modal evolution. 
In such cases, the temporal response is more naturally described by a full transfer function matrix containing cross-modal terms rather than by the purely diagonal representation of Eq.~\ref{eq:diag_weight_tf}.
This provides a likely explanation for the comparatively poor performance of analytical tGEM and the substantial improvement obtained when cross-modal interactions were incorporated through the empirical transfer function matrices.

\subsection{Cross-modal Dynamics and Empirical Transfer Functions}
Although the analytical transfer function constraint did not consistently improve source reconstruction, the empirical results indicate that temporal information can provide an effective constraint when the underlying modal dynamics are represented more accurately.
In contrast to the diagonal analytical formulation, the empirical transfer function matrix allows activity in one eigenmode to contribute to the temporal response of another. 
This relaxation of independent modal evolution produced substantial improvements in reconstruction, particularly in the presence of measurement noise. 
These results suggest that the useful temporal structure of cortical activity is not captured solely by the characteristic frequency response assigned to each individual geometric eigenmode, but also by the interactions between modes.

The performance of the empirical approaches also depended on how specifically the transfer functions represented the underlying dynamics.
Transfer functions averaged across all simulations provided improvements under noisy conditions, but generally performed worse than those estimated separately for each simulation.
This indicates that some cross-modal relationships are sufficiently consistent to provide a useful population-level constraint, while additional information is contained in simulation-specific dynamics. 
The hybrid approaches showed a similar pattern and generally provided the strongest non-oracle performance. 
This may reflect the complementary information contained in the analytical and empirical components. 
The empirical transfer function matrix captures cross-modal relationships, but its normalisation results in unity along the diagonal and thus removes the mode-specific scaling of the self-transfer terms. 
The hybrid formulation reintroduces this scaling using the analytical neural-field transfer function while retaining the empirically estimated cross-modal structure.
Therefore, its improved performance suggests that, although the analytical transfer function alone is too restrictive, its mode- and frequency-dependent scaling may remain informative when combined with a more flexible representation of cross-modal dynamics.

The oracle results should be interpreted cautiously, since improved performance is expected when the transfer function matrix is estimated from the true underlying source dynamics. 
Rather than representing a practical reconstruction method, the oracle thus provides an upper bound on the benefit obtainable from perfectly specified time-varying cross-modal information \cite{vincent_oracle_2007}.
However, its strong performance shows that a transfer function averaged over an entire seizure cannot fully capture the changing relationship between spatial modes as the system transitions between dynamical regimes. 
Taken together, these findings suggest that the principal challenge, for temporal constraining of the source localisation problem in eigenmode space, is how accurately the evolving cross-modal dynamics can be estimated without access to the underlying source activity.

\subsection{Estimating the Cross-Modal Coupling Terms in Practice}
The strong performance of the empirical and oracle approaches raises an important practical question: how could the required transfer function matrices be estimated without knowledge of the true underlying source activity? 
In the present simulation study, the source dynamics were known and could therefore be projected directly into the eigenmode basis to estimate the corresponding cross-modal relationships. 
However, with experimental scalp EEG, the underlying source activity is itself unknown. 
We thus have a seemingly recursive problem in that an accurate source estimate is required to estimate the transfer functions to produce an accurate source estimate.

One practical way to break this chicken-and-egg problem is to estimate the relevant dynamical relationships using complementary measurements that provide more direct information about the underlying neural activity.
Simultaneous or previously acquired intracranial recordings are particularly attractive in this context because they provide substantially greater spatial specificity than scalp EEG, while retaining temporal resolution \cite{mikulan_simultaneous_2020, parmigiani_simultaneous_2022}. 
Related approaches have been used to constrain neural field model parameters from intracranial data.
For example, \textcite{phogat_unified_2025} estimated coupling and gain parameters from invasive recordings and showed that the resulting model could reproduce features of the measured dynamics. 
Such measurements could similarly be used to estimate cross-modal transfer relationships, which could subsequently be applied as priors during electrophysiological source localisation as performed in this paper. 
Functional MRI provides another potential source of information.
Geometric eigenmodes have already been used extensively to characterise spatial patterns of fMRI activity \cite{pang_geometric_2023, li_mapping_2025, potash_investigating_2025}, making fMRI a natural candidate for estimating relationships between spatial modes. 
However, its substantially lower temporal resolution, prevents direct estimation of the frequency-dependent electrophysiological transfer functions considered here \cite{glover_overview_2011}. 
A possible alternative would be a hybrid formulation in which fMRI is used to estimate approximately frequency-independent cross-modal coupling terms, while the frequency-dependent scaling of individual modes is retained from the analytical neural-field transfer function. 
This would be conceptually similar to the hybrid approach considered in the present study, combining empirically estimated interactions between modes with analytically prescribed mode-specific temporal dynamics.

An alternative would be to derive a generalised analytical transfer function model that explicitly incorporates coupling between eigenmodes. 
In principle, this would remove the need to estimate the cross-modal terms directly from source activity. 
In practice, however, such a derivation is considerably more mathematically substantial than the diagonal formulation used in this study, and would strongly depend on assumptions about the coupling architecture, population dynamics, and operating regime of the underlying neural-field model \cite{gabay_dynamics_2018}.
Therefore, the resulting analytical model would introduce additional physiological and mathematical assumptions that would themselves require validation.

The present study does not attempt to solve this estimation problem. 
Instead, we address the more fundamental question of whether empirically informed transfer functions would provide useful constraints if such information were available and how their utility depends on the level of specificity at which they are estimated. 
Therefore, we considered three progressively more informative cases: the averaged transfer function represents a population-level prior shared across simulations; the simulation-specific transfer function approximates a condition- or subject-specific prior that is fixed over time; and the oracle transfer function represents the limiting case in which the transfer matrix is known locally in time. 
These cases provide a hierarchy for assessing how increasingly specific knowledge of the underlying dynamics affects source reconstruction quality.

\paragraph{Population-level transfer functions}
The simplest practical implementation would be to learn a common transfer function matrix from a large calibration dataset and apply it to previously unseen recordings.
This corresponds most closely to the averaged empirical formulation tested here, in which a single transfer function matrix was shared across all simulations.
Large multimodal neuroimaging datasets containing electrophysiological, anatomical, and behavioural information provide a potential basis for identifying modal relationships that are sufficiently conserved across individuals \cite{van_essen_wu-minn_2013, sun_harvard_2025}.
Once estimated, such a population-level transfer function could be incorporated as a fixed prior with little additional calibration required for a new subject. 
The improved performance of the averaged empirical approaches under noisy conditions suggests that useful dynamical structure can survive this level of averaging. 
However, their poorer performance relative to simulation-specific estimates also indicates that a single population-level matrix necessarily discards variability associated with the individual.

\paragraph{Condition-specific and subject-specific transfer functions}
A more informative prior could be obtained by restricting the calibration data to recordings that more closely match the subject or condition being analysed. 
In epilepsy, for example, transfer functions could be estimated separately according to seizure type, anatomical onset region, or stage of seizure evolution. 
This is motivated by the substantial changes in seizure dynamics that occur between onset, propagation, and termination, and across different dynamical seizure classes \cite{jirsa_nature_2014, saggio_taxonomy_2020}. 
More generally, the empirical transfer function matrix could be conditioned on a state variable $c$, giving $H^{\mathrm{E}}_{ji}(f|c)$ rather than assuming the same modal relationships across all recordings.

Subject-specific calibration provides a further level of specificity. 
For patients undergoing repeated or long-term monitoring, as may be undertaken in chronic conditions such as epilepsy \cite{halliday_umpire_2025, sanger_gamma_2026}, empirical transfer functions could, in principle, be estimated from previous recordings and subsequently reused as priors for later non-invasive source reconstruction. 
Simultaneous measurements with complementary modalities could be particularly useful for this calibration. 
Multimodal constraints are already widely used in EEG source imaging \cite{lei_incorporating_2015}, from intracranial EEG \cite{hosseini_electromagnetic_2018}, MEG \cite{molins_quantification_2008}, and fMRI \cite{he_multimodal_2008}.  
However, the principal limitation is that these alternate modalities typically operate at different spatial and temporal resolutions of EEG, meaning that they may be insufficient for capturing the spatially distributed sources and modes of activity in electrophysiological source imaging. 

\paragraph{Time-varying transfer functions}
The most informative but also most difficult extension would be to allow the empirical transfer function to change continuously with the underlying dynamical state. This corresponds to the oracle formulation considered here, in which the transfer function matrix was estimated separately for each time window. This approach is clearly not available directly in experimental EEG since the source dynamics required to estimate $H^{\mathrm{E}}_{ji}(f,\tau)$ are precisely the quantities being inferred. Nevertheless, the oracle results provide a useful indication of the potential benefit of tracking nonstationary modal relationships rather than assuming that a single transfer function applies throughout a recording.

A practical approximation may, therefore, be to update the transfer function prior only when the system enters a different dynamical regime, rather than attempting to estimate a new matrix at every STFT window.
For seizures, this could involve switching between transfer functions associated with onset, propagation, and termination, provided these states can be identified reliably from the measured data. 
Such a formulation would sit between the stationary condition-specific estimates and the fully time-resolved oracle case, while requiring substantially less information than the latter.

\subsection{Conclusion}
Overall, this study shows that temporal constraints can improve geometric-eigenmode source localisation, but only when the underlying modal dynamics are appropriately captured.
Analytical neural field transfer functions did not consistently improve reconstruction, whereas empirically informed cross-modal transfer functions were substantially more effective, particularly under noisy conditions. 
Even a stationary population level estimate can provide useful temporal constraints, while increasingly specific representations of the underlying dynamics progressively approach the reconstruction performance of the oracle case. 
These findings suggest that improving the representation of interactions between modes is an important next step for both geometric eigenmodes as spatial constraints and neural field modelling more broadly.

\appendix

% [[! May need some rewriting...]]
\section{Structural Brain Data \label{app:sim_structural_data}}
Structural brain data from the Human Connectome Project (HCP) \cite{van_essen_wu-minn_2013} was used, with HCP Subject 100206 (HCP100206) chosen as the primary subject for this work.
The T1 MRI image with the skull included was used to generate surface meshes in the subject's native space for the skin, skull, and cortex using recon-all from \textsc{Freesurfer} \cite{fischl_freesurfer_2012}.
The cortical mesh from fsaverage (\textsc{Freesurfer}'s default) template space was also used to test the effects of using group-averaged structural data in the source imaging pipeline. 
The individual subject's cortical mesh was aligned to the fsaverage template mesh, which consisted of 163,842 vertices per hemisphere.

\subsubsection{Parcellation Atlas}
Each vertex was attributed to one of the cortical regions of the Yan 1000 region homotopic atlas \cite{yan_homotopic_2023}.
The location of each cortical region was defined as the centroid of its vertex coordinates.
In addition to the 1000 cortical regions, 19 subcortical structures defined by the cifti grayordinate template \cite{glasser_minimal_2013} were also included, resulting in a 1019-region parcellation for the entire brain.   

\subsubsection{Structural Connectome}
The individual’s structural connectome was computed by a collaborator via a diffusion MRI tractography pipeline detailed elsewhere \cite{mansourl_high-resolution_2021, mansour_l_connectomes_2023} using the MRtrix3 software \cite{tournier_mrtrix3_2019}. 
The fibre orientation distributions (FODs) were computed from tissue-specific response functions via Multi-Shell Multi-Tissue Constrained Spherical Deconvolution \cite{jeurissen_multi-tissue_2014, dhollander_unsupervised_2016}. 
A total of 5 million tractography streamlines were estimated by an anatomically constrained probabilistic tractography algorithm employing second-order integration over FODs \cite{tournier_improved_2010, smith_anatomically-constrained_2012}.
Notably, streamlines were randomly seeded from the gray matter white matter interface. 
A radial search with a maximum radius threshold of 4~mm was used to map streamlines to the regional parcellation, resulting in a 1019x1019 connectivity matrix of streamline counts. 
The matrix of streamline counts was then normalised by the maximum streamline count to end in a normalised connectivity matrix with entries between 0 and 1.  
In addition, a 1019x1019 streamline length matrix was computed to quantify average streamline lengths between all pairs of connected regions.
To obtain the group-averaged connectivity matrix of streamline counts, the sum of streamlines between regions was taken for every patient, and the aggregate matrix is normalised between 0 and 1.  

\subsubsection{Dipole Source Space}

The dipole source space is a representation of all dipole neural sources, located on the cortical surface, which can be summed to generate EEG/MEG data. 
These dipoles are nominally oriented normal to the cortical surface, but some flexibility can be allowed for the orientation to be `loose' and deviate slightly from the normal. 
The location of the dipole sources were defined by subdividing an icosahedron and mapping the resulting vertices onto the cortical surface.
Depending on the subdivision level, the number of dipoles ranged from 1,284 to 20,484, with higher densities reducing errors associated with overly large dipoles that do not adequately capture cortical geometry, at the cost of increased computational complexity.
In this framework, seizure simulations were mapped to the source space using a one-to-many correspondence based on the cortical parcellation, such that each cortical region in the Yan-1000 atlas contributed the same regional activity to all dipoles associated with that region.
To mitigate boundary artefacts introduced during this projection, Laplacian smoothing was applied across neighbouring dipoles for five iterations.

\section{Epileptor Model \label{app:sim_epileptor}}

The Epileptor model is a phenomenological dynamical model that has remarkable success in simulating seizure-like activity \cite{jirsa_nature_2014}.  

\subsection{Epileptor Equations}
The most common implementation of the Epileptor model, as implemented in the Virtual Brain Project (\textsc{TVB}) \cite{sanz_leon_virtual_2013}, characterises the dynamical behaviour of epileptic seizures using five state variables $(x_1,y_1,x_2,y_2,z)$ plus a dummy variable $(u)$ within region $i$.  
The Epileptor equations are
\begin{equation}
\label{eq:6df_Epileptor}
\begin{split}
    \tau_1 \dot x_1 &= y_1 - f_1(x_1,x_2,z) - z + I_{rest,1},\\
    \dot y_1 &= y_0 - 5 x_1^2 - y_1,\\
    \tau_0  \dot z &= 4(x_1- x_0) - z + J,\\
    \dot x_2 &= -y_2 + x_2 - x_2^3 + I_{rest,2} + 2u - 0.3(z-3.5),\\
    \tau_2 \dot y_2 &= -y_2 + f_2(x_2),\\
    \dot u &= -\gamma (u - 0.1 x_1),
\end{split}
\end{equation}
where
\begin{equation*}
\begin{split}
f_1(x_1,x_2,z) &= \begin{cases} \mbox{$x_1^3 - 3 x_1^2$} & \mbox{if } x_1 < 0 \\ \mbox{($x_2 - 0.6 (z-4)^2)  x_1$} & \mbox{otherwise} \end{cases} , \\
f_2(x_2) &= \begin{cases} \mbox{$0$} & \mbox{if } x_2 < -0.25 \\ \mbox{$6(x_2 + 0.25)$} & \mbox{otherwise} \end{cases} , \\
%u &= 0.001g(x_1) = 0.001\int^t_{t_0} e^{-\gamma(t-s)} x_1(s) ds ,\\
\tau_{ij} &= \frac{D_{ij}}{v}, \\ 
J &= - g \sum_{j=1}^N W_{ij} [x_{1,j}(t-\tau_{ij}) - x_{1,i}(t)].
\end{split}
\end{equation*}

\begin{table}[htbp]
\centering
\bcaption{Table of variables and parameters of the Epileptor}{}
\label{tab:Epileptor_params}
\begin{tabular*}{\linewidth}{@{}ccp{300pt}@{}}
\toprule
\multicolumn{2}{c}{State variables}    & Interpretation        \\ \midrule
\multicolumn{2}{c}{$x_1(t), y_1(t)$}   & Fast timescale rapid discharges \\
\multicolumn{2}{c}{$x_2(t), y_2(t)$}   & Intermediate timescale spike-and-waves\\
\multicolumn{2}{c}{$z(t)$} & Slow timescale transition between seizure and non-seizure \\
\multicolumn{2}{c}{$u(t)$} & Dummy variable for low-pass filtering from $x_1$ to $x_2$ \\  \midrule
\multicolumn{2}{c}{Adjusted parameters}&               \\ \midrule
$x_0$       & $-2.1 \rightarrow -1.6$  & Node excitability \\
$g$         & $1.5 \rightarrow 2.0$    & Global coupling strength \\
$W_{ij}$    & $0.0 \rightarrow 1$      & Coupling weights defined by 
SCM into region $i$ from $j$ \\  
$D_{ij}$    & $0.0 \rightarrow 248.5$  & Streamline lengths between regions (mm) \\
$\tau_{ij}$ & $0.8 \rightarrow 82.8$  & Time delays between regions (ms) \\  \midrule

\multicolumn{2}{c}{Fixed parameters}       &               \\ \midrule
$I_{rest,1}$& 3.1                      & Constant injection current for $x_1$\\ 
$I_{rest,2}$& 0.45                     & Constant injection current for $x_1$\\ 
$y_0$       & 1                        & Threshold constant for $y_1$ \\ 
$\gamma$    & 0.01                     & Time constant of low-pass filter \\ 
$\tau_1$    & 1                        & Time constant of of fast activity\\
$\tau_2$    & 10                       & Time constant of spike-and-waves\\ 
$r=1/\tau_0$& $0.00015$  & Temporal scaling of long timescale activity \\ 
$v$         & 3                        & Axon conduction velocity ($\mathrm{ms^{-1}}$) \\
$\sigma^2$  & $0.0001$& Stochastic noise term (var) affecting frequency of inter-ictal spikes and seizures \\ 
\bottomrule
\end{tabular*}
\end{table}

\subsubsection*{Representation as Dipoles}
The Epileptor model generates time series representing neural activity within a single region of the brain.  
To represent extracellular dipoles ($x$), the Epileptor equations are modified to become
\begin{equation} \label{eq:epileptor_source}
    x = (x_1 - x_2 - b) /n \times10^{-5},
\end{equation}
where $x_1 - x_2$ is used instead of $x_2-x_1$, as in \textcite{jirsa_nature_2014}, to ensure that the seizure state has a positive orientation, $n$ is the number of dipole sources in the source space, and $10^{-5}$ converts to units of $\mathrm{Am}$.  
Notably, $b$ is a baseline constant, subtracted from the Epileptor model in its original form, to ensure dipole sources maintain approximately zero baseline activity, as expected from the Debye shielding of ions in the electrolyte solution of the extracellular space \cite{nunez_electric_2006}.  

\subsubsection*{Coupled Neural Mass Models for Seizure Spread}
To generate seizure activity across the entire brain, simulations involved coupled masses of Epileptor models, as defined in Eq.~\ref{eq:6df_Epileptor}, through the coupling current $J$. 
In such a framework, each neural mass model corresponds to one cortical or subcortical region as per the Yan-1000 cortical parcellation atlas \cite{yan_homotopic_2023}.
The connectivity weight matrix ($W_{ij}$) is henceforth defined by the structural connectome (see Sec.~\ref{app:sim_structural_data}) of the subject.
The axonal time delay $\tau_{ij}$ was defined by the 1019 region connectome streamline lengths ($D_{ij}$) divided by a conduction speed of 3 m/s \cite{sanz_leon_virtual_2013}. 

A Heun's stochastic integrator was used with Gaussian additive noise ($\sim N(0, \sigma^2)$) added to the variables $x_2$ and $y_2$  \cite{jirsa_nature_2014, moosavi_critical_2022}.
An integration step-size $\Delta t = 0.05$ was used.  
To match the temporal properties of focal seizures, the timescale of the model output was scaled such that 256 time steps correspond to 1 s.
Seizure simulations were chosen to last one minute to enable sufficient epileptogenic activity to develop.
To generate each simulation, a single given region was first defined as the \lq{}epileptogenic zone\rq{}. All regions other than the epileptogenic zone were set to have an excitability of $x_0 = -2.1$; i.e., tuned to be at the cusp of a bifurcation \cite{houssaini_epileptor_2020, moosavi_critical_2022}.
To trigger a seizure, the excitability ($x_0$) for the specified epileptogenic zone was increased to $-1.6$.  

\section{Metrics \label{app:metrics}}
To assess the accuracy of source localisation within simulations, the following evaluation metrics were used:

\paragraph{Cosine Score $S_C$}
\begin{equation}
    S_C = \frac{\hat{x}\cdot x}{||\hat{x}||||x||} ,   
\end{equation}
where $x$ is the true signal and $\hat{x}$ is the source estimate.
Dependent on the figure, $S_C$ was presented for each time point or averaged across time.
The cosine score captures relative distributional accuracy, reflecting how well the relative pattern of activation across the source space matches the true activity.

\paragraph{Normalised Residual Sum of Squares $nRSS$} 
\begin{equation}
    \mathit{nRSS} = \sum_i \left( \frac{\hat{x}}{\text{max}(|\hat{x}|)} - \frac{x}{\text{max}(|x|)}\right)^2,
\end{equation}
where the data was normalised using the maximal value, to compare across methods with differing units.

\paragraph{Mean Squared Error $MSE$} 
\begin{equation}
    \mathit{MSE} = \frac{1}{N}\sum_i \left( \frac{\hat{x}}{\text{max}(|\hat{x}|)} - \frac{x}{\text{max}(|x|)}\right)^2 = \frac{nRSS}{N},
\end{equation}
where $N=1000$ is the number of dipole sources.

\paragraph{Variance Explained $R^2$}
\begin{equation}
    \mathit{R^2} = 1 - \frac{\Sigma_i(\hat{x}-x)^2}{\Sigma_ix^2} = 1 - \frac{nRSS}{nSS},
\end{equation}
where $nRSS$ is the normalised residual sum of squares, and $nSS$ is the normalised sum of squares over all possible sources, $i$, in source space.
Note that for the simulated signals used here, a baseline constant was already identified and subtracted (see Equation~\ref{eq:epileptor_source}) such that the baseline for $x$ is at 0. 

\paragraph{Region Localisation Error $RLE$ }
\begin{equation}
    \mathit{RLE} = \frac{1}{2Q}\sum_{k \in I} \min_{l \in \hat{I}}{||r_k - r_l||} + \frac{1}{2\hat{Q}}\sum_{l \in \hat{I}} \min_{k \in I}{||r_k - r_l||},
\end{equation}
where $I$ and $\hat{I}$ represent the true and estimated indices of active sources, respectively, $Q$ and $\hat{Q}$ represent the number of true and estimated active sources, respectively, and $r_k$ denotes the position of the $k$-th dipole source in space as per \textsc{MNE-Python}'s implementation \cite{gramfort_mne_2014}.
Sufficiently active sources were algorithmically identified to be those beyond Otsu's Threshold \cite{otsu_threshold_1979} as suggested by \textcite{sun_seizure_2024}. 
The region localisation error was computed at $50\%$ of maximum EEG power, representing a sensible extent of seizure spread and the rising phase of a seizure \cite{plummer_interictal_2019}.
Region localisation error captures spatial accuracy, reflecting how precisely the estimated source activity is positioned in space.

\section{Data Availability Statement}
The data that support the findings of this study were generated by numerical simulations. 
The source code and parameters used to generate the simulations is publicly available \cite{siu_github_sim, siu_neural_2026}.
The algorithms used in this study are also publicly available for others to apply to their own data and analyses \cite{siu_github_sesl}.

\begin{acknowledgments}
The authors thank Dr. Richa Phogat for thoughtful discussions on neural field theory and her published open-source code.
This research was supported by The University of Melbourne’s Research Computing Services, the Petascale Campus Initiative, and an Australian Government Research Training Program (RTP) Scholarship.
This research was funded by the Australian Research Council (Discovery Project DP200102600).
\end{acknowledgments}

\bibliography{references}

\begin{thebibliography}{60}%
\makeatletter
\providecommand \@ifxundefined [1]{%
 \@ifx{#1\undefined}
}%
\providecommand \@ifnum [1]{%
 \ifnum #1\expandafter \@firstoftwo
 \else \expandafter \@secondoftwo
 \fi
}%
\providecommand \@ifx [1]{%
 \ifx #1\expandafter \@firstoftwo
 \else \expandafter \@secondoftwo
 \fi
}%
\providecommand \natexlab [1]{#1}%
\providecommand \enquote  [1]{``#1''}%
\providecommand \bibnamefont  [1]{#1}%
\providecommand \bibfnamefont [1]{#1}%
\providecommand \citenamefont [1]{#1}%
\providecommand \href@noop [0]{\@secondoftwo}%
\providecommand \href [0]{\begingroup \@sanitize@url \@href}%
\providecommand \@href[1]{\@@startlink{#1}\@@href}%
\providecommand \@@href[1]{\endgroup#1\@@endlink}%
\providecommand \@sanitize@url [0]{\catcode `\\12\catcode `\$12\catcode `\&12\catcode `\#12\catcode `\^12\catcode `\_12\catcode `\%12\relax}%
\providecommand \@@startlink[1]{}%
\providecommand \@@endlink[0]{}%
\providecommand \url  [0]{\begingroup\@sanitize@url \@url }%
\providecommand \@url [1]{\endgroup\@href {#1}{\urlprefix }}%
\providecommand \urlprefix  [0]{URL }%
\providecommand \Eprint [0]{\href }%
\providecommand \doibase [0]{https://doi.org/}%
\providecommand \selectlanguage [0]{\@gobble}%
\providecommand \bibinfo  [0]{\@secondoftwo}%
\providecommand \bibfield  [0]{\@secondoftwo}%
\providecommand \translation [1]{[#1]}%
\providecommand \BibitemOpen [0]{}%
\providecommand \bibitemStop [0]{}%
\providecommand \bibitemNoStop [0]{.\EOS\space}%
\providecommand \EOS [0]{\spacefactor3000\relax}%
\providecommand \BibitemShut  [1]{\csname bibitem#1\endcsname}%
\let\auto@bib@innerbib\@empty
%</preamble>
\bibitem [{\citenamefont {Breakspear}(2017)}]{breakspear_dynamic_2017}%
  \BibitemOpen
  \bibfield  {author} {\bibinfo {author} {\bibfnamefont {M.}~\bibnamefont {Breakspear}},\ }\href {https://doi.org/10.1038/nn.4497} {\bibfield  {journal} {\bibinfo  {journal} {Nature Neuroscience}\ }\textbf {\bibinfo {volume} {20}},\ \bibinfo {pages} {340} (\bibinfo {year} {2017})}\BibitemShut {NoStop}%
\bibitem [{\citenamefont {Deco}\ \emph {et~al.}(2008)\citenamefont {Deco}, \citenamefont {Jirsa}, \citenamefont {Robinson}, \citenamefont {Breakspear},\ and\ \citenamefont {Friston}}]{deco_dynamic_2008}%
  \BibitemOpen
  \bibfield  {author} {\bibinfo {author} {\bibfnamefont {G.}~\bibnamefont {Deco}}, \bibinfo {author} {\bibfnamefont {V.~K.}\ \bibnamefont {Jirsa}}, \bibinfo {author} {\bibfnamefont {P.~A.}\ \bibnamefont {Robinson}}, \bibinfo {author} {\bibfnamefont {M.}~\bibnamefont {Breakspear}},\ and\ \bibinfo {author} {\bibfnamefont {K.}~\bibnamefont {Friston}},\ }\href {https://doi.org/10.1371/journal.pcbi.1000092} {\bibfield  {journal} {\bibinfo  {journal} {PLoS Computational Biology}\ }\textbf {\bibinfo {volume} {4}},\ \bibinfo {pages} {e1000092} (\bibinfo {year} {2008})}\BibitemShut {NoStop}%
\bibitem [{\citenamefont {Jirsa}\ and\ \citenamefont {Haken}(1996)}]{jirsa_field_1996}%
  \BibitemOpen
  \bibfield  {author} {\bibinfo {author} {\bibfnamefont {V.~K.}\ \bibnamefont {Jirsa}}\ and\ \bibinfo {author} {\bibfnamefont {H.}~\bibnamefont {Haken}},\ }\href {https://doi.org/10.1103/PhysRevLett.77.960} {\bibfield  {journal} {\bibinfo  {journal} {Physical Review Letters}\ }\textbf {\bibinfo {volume} {77}},\ \bibinfo {pages} {960} (\bibinfo {year} {1996})}\BibitemShut {NoStop}%
\bibitem [{\citenamefont {Gabay}\ and\ \citenamefont {Robinson}(2017)}]{gabay_cortical_2017}%
  \BibitemOpen
  \bibfield  {author} {\bibinfo {author} {\bibfnamefont {N.~C.}\ \bibnamefont {Gabay}}\ and\ \bibinfo {author} {\bibfnamefont {P.~A.}\ \bibnamefont {Robinson}},\ }\href {https://doi.org/10.1103/PhysRevE.96.032413} {\bibfield  {journal} {\bibinfo  {journal} {Physical Review E}\ }\textbf {\bibinfo {volume} {96}},\ \bibinfo {pages} {032413} (\bibinfo {year} {2017})}\BibitemShut {NoStop}%
\bibitem [{\citenamefont {Pang}\ \emph {et~al.}(2023)\citenamefont {Pang}, \citenamefont {Aquino}, \citenamefont {Oldehinkel}, \citenamefont {Robinson}, \citenamefont {Fulcher}, \citenamefont {Breakspear},\ and\ \citenamefont {Fornito}}]{pang_geometric_2023}%
  \BibitemOpen
  \bibfield  {author} {\bibinfo {author} {\bibfnamefont {J.~C.}\ \bibnamefont {Pang}}, \bibinfo {author} {\bibfnamefont {K.~M.}\ \bibnamefont {Aquino}}, \bibinfo {author} {\bibfnamefont {M.}~\bibnamefont {Oldehinkel}}, \bibinfo {author} {\bibfnamefont {P.~A.}\ \bibnamefont {Robinson}}, \bibinfo {author} {\bibfnamefont {B.~D.}\ \bibnamefont {Fulcher}}, \bibinfo {author} {\bibfnamefont {M.}~\bibnamefont {Breakspear}},\ and\ \bibinfo {author} {\bibfnamefont {A.}~\bibnamefont {Fornito}},\ }\href {https://doi.org/10.1038/s41586-023-06098-1} {\bibfield  {journal} {\bibinfo  {journal} {Nature}\ ,\ \bibinfo {pages} {1}} (\bibinfo {year} {2023})}\BibitemShut {NoStop}%
\bibitem [{\citenamefont {Robinson}\ \emph {et~al.}(2016)\citenamefont {Robinson}, \citenamefont {Zhao}, \citenamefont {Aquino}, \citenamefont {Griffiths}, \citenamefont {Sarkar},\ and\ \citenamefont {Mehta-Pandejee}}]{robinson_eigenmodes_2016}%
  \BibitemOpen
  \bibfield  {author} {\bibinfo {author} {\bibfnamefont {P.~A.}\ \bibnamefont {Robinson}}, \bibinfo {author} {\bibfnamefont {X.}~\bibnamefont {Zhao}}, \bibinfo {author} {\bibfnamefont {K.~M.}\ \bibnamefont {Aquino}}, \bibinfo {author} {\bibfnamefont {J.~D.}\ \bibnamefont {Griffiths}}, \bibinfo {author} {\bibfnamefont {S.}~\bibnamefont {Sarkar}},\ and\ \bibinfo {author} {\bibfnamefont {G.}~\bibnamefont {Mehta-Pandejee}},\ }\href {https://doi.org/10.1016/j.neuroimage.2016.04.050} {\bibfield  {journal} {\bibinfo  {journal} {NeuroImage}\ }\textbf {\bibinfo {volume} {142}},\ \bibinfo {pages} {79} (\bibinfo {year} {2016})}\BibitemShut {NoStop}%
\bibitem [{\citenamefont {Gabay}\ \emph {et~al.}(2018)\citenamefont {Gabay}, \citenamefont {Babaie-Janvier},\ and\ \citenamefont {Robinson}}]{gabay_dynamics_2018}%
  \BibitemOpen
  \bibfield  {author} {\bibinfo {author} {\bibfnamefont {N.~C.}\ \bibnamefont {Gabay}}, \bibinfo {author} {\bibfnamefont {T.}~\bibnamefont {Babaie-Janvier}},\ and\ \bibinfo {author} {\bibfnamefont {P.~A.}\ \bibnamefont {Robinson}},\ }\href {https://doi.org/10.1103/PhysRevE.98.042413} {\bibfield  {journal} {\bibinfo  {journal} {Physical Review E}\ }\textbf {\bibinfo {volume} {98}},\ \bibinfo {pages} {042413} (\bibinfo {year} {2018})}\BibitemShut {NoStop}%
\bibitem [{\citenamefont {He}\ \emph {et~al.}(2018)\citenamefont {He}, \citenamefont {Sohrabpour}, \citenamefont {Brown},\ and\ \citenamefont {Liu}}]{he_electrophysiological_2018}%
  \BibitemOpen
  \bibfield  {author} {\bibinfo {author} {\bibfnamefont {B.}~\bibnamefont {He}}, \bibinfo {author} {\bibfnamefont {A.}~\bibnamefont {Sohrabpour}}, \bibinfo {author} {\bibfnamefont {E.}~\bibnamefont {Brown}},\ and\ \bibinfo {author} {\bibfnamefont {Z.}~\bibnamefont {Liu}},\ }\href {https://doi.org/10.1146/annurev-bioeng-062117-120853} {\bibfield  {journal} {\bibinfo  {journal} {Annual Review of Biomedical Engineering}\ }\textbf {\bibinfo {volume} {20}},\ \bibinfo {pages} {171} (\bibinfo {year} {2018})}\BibitemShut {NoStop}%
\bibitem [{\citenamefont {Hallez}\ \emph {et~al.}(2007)\citenamefont {Hallez}, \citenamefont {Vanrumste}, \citenamefont {Grech}, \citenamefont {Muscat}, \citenamefont {De~Clercq}, \citenamefont {Vergult}, \citenamefont {D'Asseler}, \citenamefont {Camilleri}, \citenamefont {Fabri}, \citenamefont {Van~Huffel},\ and\ \citenamefont {Lemahieu}}]{hallez_review_2007}%
  \BibitemOpen
  \bibfield  {author} {\bibinfo {author} {\bibfnamefont {H.}~\bibnamefont {Hallez}}, \bibinfo {author} {\bibfnamefont {B.}~\bibnamefont {Vanrumste}}, \bibinfo {author} {\bibfnamefont {R.}~\bibnamefont {Grech}}, \bibinfo {author} {\bibfnamefont {J.}~\bibnamefont {Muscat}}, \bibinfo {author} {\bibfnamefont {W.}~\bibnamefont {De~Clercq}}, \bibinfo {author} {\bibfnamefont {A.}~\bibnamefont {Vergult}}, \bibinfo {author} {\bibfnamefont {Y.}~\bibnamefont {D'Asseler}}, \bibinfo {author} {\bibfnamefont {K.~P.}\ \bibnamefont {Camilleri}}, \bibinfo {author} {\bibfnamefont {S.~G.}\ \bibnamefont {Fabri}}, \bibinfo {author} {\bibfnamefont {S.}~\bibnamefont {Van~Huffel}},\ and\ \bibinfo {author} {\bibfnamefont {I.}~\bibnamefont {Lemahieu}},\ }\href {https://doi.org/10.1186/1743-0003-4-46} {\bibfield  {journal} {\bibinfo  {journal} {Journal of NeuroEngineering and Rehabilitation}\ }\textbf {\bibinfo {volume} {4}},\ \bibinfo {pages} {46} (\bibinfo {year} {2007})}\BibitemShut {NoStop}%
\bibitem [{\citenamefont {Hauk}\ \emph {et~al.}(2022)\citenamefont {Hauk}, \citenamefont {Stenroos},\ and\ \citenamefont {Treder}}]{hauk_towards_2022}%
  \BibitemOpen
  \bibfield  {author} {\bibinfo {author} {\bibfnamefont {O.}~\bibnamefont {Hauk}}, \bibinfo {author} {\bibfnamefont {M.}~\bibnamefont {Stenroos}},\ and\ \bibinfo {author} {\bibfnamefont {M.~S.}\ \bibnamefont {Treder}},\ }\href {https://doi.org/10.1016/j.neuroimage.2022.119177} {\bibfield  {journal} {\bibinfo  {journal} {NeuroImage}\ }\textbf {\bibinfo {volume} {255}},\ \bibinfo {pages} {119177} (\bibinfo {year} {2022})}\BibitemShut {NoStop}%
\bibitem [{\citenamefont {Kincses}\ \emph {et~al.}(1999)\citenamefont {Kincses}, \citenamefont {Braun}, \citenamefont {Kaiser},\ and\ \citenamefont {Elbert}}]{kincses_modeling_1999}%
  \BibitemOpen
  \bibfield  {author} {\bibinfo {author} {\bibfnamefont {W.~E.}\ \bibnamefont {Kincses}}, \bibinfo {author} {\bibfnamefont {C.}~\bibnamefont {Braun}}, \bibinfo {author} {\bibfnamefont {S.}~\bibnamefont {Kaiser}},\ and\ \bibinfo {author} {\bibfnamefont {T.}~\bibnamefont {Elbert}},\ }\href {https://doi.org/10.1002/(SICI)1097-0193(1999)8:4<182::AID-HBM3>3.0.CO;2-M} {\bibfield  {journal} {\bibinfo  {journal} {Human Brain Mapping}\ }\textbf {\bibinfo {volume} {8}},\ \bibinfo {pages} {182} (\bibinfo {year} {1999})}\BibitemShut {NoStop}%
\bibitem [{\citenamefont {Pascual-Marqui}\ \emph {et~al.}(2002)\citenamefont {Pascual-Marqui}, \citenamefont {Esslen}, \citenamefont {Kochi},\ and\ \citenamefont {Lehmann}}]{pascual-marqui_functional_2002}%
  \BibitemOpen
  \bibfield  {author} {\bibinfo {author} {\bibfnamefont {R.~D.}\ \bibnamefont {Pascual-Marqui}}, \bibinfo {author} {\bibfnamefont {M.}~\bibnamefont {Esslen}}, \bibinfo {author} {\bibfnamefont {K.}~\bibnamefont {Kochi}},\ and\ \bibinfo {author} {\bibfnamefont {D.}~\bibnamefont {Lehmann}},\ }\href@noop {} {\bibfield  {journal} {\bibinfo  {journal} {Methods and Findings in Experimental and Clinical Pharmacology}\ }\textbf {\bibinfo {volume} {24 Suppl C}},\ \bibinfo {pages} {91} (\bibinfo {year} {2002})}\BibitemShut {NoStop}%
\bibitem [{\citenamefont {Costa}\ \emph {et~al.}(2015)\citenamefont {Costa}, \citenamefont {Batatia}, \citenamefont {Chaari},\ and\ \citenamefont {Tourneret}}]{costa_sparse_2015}%
  \BibitemOpen
  \bibfield  {author} {\bibinfo {author} {\bibfnamefont {F.}~\bibnamefont {Costa}}, \bibinfo {author} {\bibfnamefont {H.}~\bibnamefont {Batatia}}, \bibinfo {author} {\bibfnamefont {L.}~\bibnamefont {Chaari}},\ and\ \bibinfo {author} {\bibfnamefont {J.-Y.}\ \bibnamefont {Tourneret}},\ }\href {https://doi.org/10.1109/TBME.2015.2450015} {\bibfield  {journal} {\bibinfo  {journal} {IEEE Transactions on Biomedical Engineering}\ }\textbf {\bibinfo {volume} {62}},\ \bibinfo {pages} {2888} (\bibinfo {year} {2015})},\ \bibinfo {note} {conference Name: IEEE Transactions on Biomedical Engineering}\BibitemShut {NoStop}%
\bibitem [{\citenamefont {Haufe}\ \emph {et~al.}(2011)\citenamefont {Haufe}, \citenamefont {Tomioka}, \citenamefont {Dickhaus}, \citenamefont {Sannelli}, \citenamefont {Blankertz}, \citenamefont {Nolte},\ and\ \citenamefont {Müller}}]{haufe_large-scale_2011}%
  \BibitemOpen
  \bibfield  {author} {\bibinfo {author} {\bibfnamefont {S.}~\bibnamefont {Haufe}}, \bibinfo {author} {\bibfnamefont {R.}~\bibnamefont {Tomioka}}, \bibinfo {author} {\bibfnamefont {T.}~\bibnamefont {Dickhaus}}, \bibinfo {author} {\bibfnamefont {C.}~\bibnamefont {Sannelli}}, \bibinfo {author} {\bibfnamefont {B.}~\bibnamefont {Blankertz}}, \bibinfo {author} {\bibfnamefont {G.}~\bibnamefont {Nolte}},\ and\ \bibinfo {author} {\bibfnamefont {K.-R.}\ \bibnamefont {Müller}},\ }\href {https://doi.org/10.1016/j.neuroimage.2010.09.003} {\bibfield  {journal} {\bibinfo  {journal} {NeuroImage}\ }\textbf {\bibinfo {volume} {54}},\ \bibinfo {pages} {851} (\bibinfo {year} {2011})}\BibitemShut {NoStop}%
\bibitem [{\citenamefont {Siu}\ \emph {et~al.}(2026{\natexlab{a}})\citenamefont {Siu}, \citenamefont {Karoly}, \citenamefont {Mansour~L.}, \citenamefont {Soto‐Breceda}, \citenamefont {Kuhlmann}, \citenamefont {Cook},\ and\ \citenamefont {Grayden}}]{siu_structural_2026}%
  \BibitemOpen
  \bibfield  {author} {\bibinfo {author} {\bibfnamefont {P.~H.}\ \bibnamefont {Siu}}, \bibinfo {author} {\bibfnamefont {P.~J.}\ \bibnamefont {Karoly}}, \bibinfo {author} {\bibfnamefont {S.}~\bibnamefont {Mansour~L.}}, \bibinfo {author} {\bibfnamefont {A.}~\bibnamefont {Soto‐Breceda}}, \bibinfo {author} {\bibfnamefont {L.}~\bibnamefont {Kuhlmann}}, \bibinfo {author} {\bibfnamefont {M.~J.}\ \bibnamefont {Cook}},\ and\ \bibinfo {author} {\bibfnamefont {D.~B.}\ \bibnamefont {Grayden}},\ }\href {https://doi.org/10.1002/advs.202516802} {\bibfield  {journal} {\bibinfo  {journal} {Advanced Science}\ ,\ \bibinfo {pages} {e16802}} (\bibinfo {year} {2026}{\natexlab{a}})}\BibitemShut {NoStop}%
\bibitem [{\citenamefont {Wang}\ \emph {et~al.}(2026)\citenamefont {Wang}, \citenamefont {Lou}, \citenamefont {Wei}, \citenamefont {Sheng}, \citenamefont {Tang}, \citenamefont {Peng}, \citenamefont {Shen}, \citenamefont {Mei}, \citenamefont {Chen}, \citenamefont {Gu},\ and\ \citenamefont {Liu}}]{wang_geometry_2026}%
  \BibitemOpen
  \bibfield  {author} {\bibinfo {author} {\bibfnamefont {S.}~\bibnamefont {Wang}}, \bibinfo {author} {\bibfnamefont {K.}~\bibnamefont {Lou}}, \bibinfo {author} {\bibfnamefont {C.}~\bibnamefont {Wei}}, \bibinfo {author} {\bibfnamefont {Z.}~\bibnamefont {Sheng}}, \bibinfo {author} {\bibfnamefont {J.}~\bibnamefont {Tang}}, \bibinfo {author} {\bibfnamefont {K.}~\bibnamefont {Peng}}, \bibinfo {author} {\bibfnamefont {X.}~\bibnamefont {Shen}}, \bibinfo {author} {\bibfnamefont {S.}~\bibnamefont {Mei}}, \bibinfo {author} {\bibfnamefont {L.}~\bibnamefont {Chen}}, \bibinfo {author} {\bibfnamefont {D.}~\bibnamefont {Gu}},\ and\ \bibinfo {author} {\bibfnamefont {Q.}~\bibnamefont {Liu}},\ }\href {https://doi.org/10.1038/s41551-026-01664-0} {\bibfield  {journal} {\bibinfo  {journal} {Nature Biomedical Engineering}\ ,\ \bibinfo {pages} {1}} (\bibinfo {year} {2026})}\BibitemShut {NoStop}%
\bibitem [{\citenamefont {Rué‐Queralt}\ \emph {et~al.}(2024)\citenamefont {Rué‐Queralt}, \citenamefont {Fluhr}, \citenamefont {Tourbier}, \citenamefont {Aleman‐Gómez}, \citenamefont {Pascucci}, \citenamefont {Yerly}, \citenamefont {Glomb}, \citenamefont {Plomp},\ and\ \citenamefont {Hagmann}}]{ruequeralt_connectome_2024}%
  \BibitemOpen
  \bibfield  {author} {\bibinfo {author} {\bibfnamefont {J.}~\bibnamefont {Rué‐Queralt}}, \bibinfo {author} {\bibfnamefont {H.}~\bibnamefont {Fluhr}}, \bibinfo {author} {\bibfnamefont {S.}~\bibnamefont {Tourbier}}, \bibinfo {author} {\bibfnamefont {Y.}~\bibnamefont {Aleman‐Gómez}}, \bibinfo {author} {\bibfnamefont {D.}~\bibnamefont {Pascucci}}, \bibinfo {author} {\bibfnamefont {J.}~\bibnamefont {Yerly}}, \bibinfo {author} {\bibfnamefont {K.}~\bibnamefont {Glomb}}, \bibinfo {author} {\bibfnamefont {G.}~\bibnamefont {Plomp}},\ and\ \bibinfo {author} {\bibfnamefont {P.}~\bibnamefont {Hagmann}},\ }\href {https://doi.org/10.1002/hbm.26638} {\bibfield  {journal} {\bibinfo  {journal} {Human Brain Mapping}\ }\textbf {\bibinfo {volume} {45}},\ \bibinfo {pages} {e26638} (\bibinfo {year} {2024})}\BibitemShut {NoStop}%
\bibitem [{\citenamefont {Robinson}\ \emph {et~al.}(2001)\citenamefont {Robinson}, \citenamefont {Rennie}, \citenamefont {Wright}, \citenamefont {Bahramali}, \citenamefont {Gordon},\ and\ \citenamefont {Rowe}}]{robinson_prediction_2001}%
  \BibitemOpen
  \bibfield  {author} {\bibinfo {author} {\bibfnamefont {P.~A.}\ \bibnamefont {Robinson}}, \bibinfo {author} {\bibfnamefont {C.~J.}\ \bibnamefont {Rennie}}, \bibinfo {author} {\bibfnamefont {J.~J.}\ \bibnamefont {Wright}}, \bibinfo {author} {\bibfnamefont {H.}~\bibnamefont {Bahramali}}, \bibinfo {author} {\bibfnamefont {E.}~\bibnamefont {Gordon}},\ and\ \bibinfo {author} {\bibfnamefont {D.~L.}\ \bibnamefont {Rowe}},\ }\href {https://doi.org/10.1103/PhysRevE.63.021903} {\bibfield  {journal} {\bibinfo  {journal} {Physical Review. E, Statistical, Nonlinear, and Soft Matter Physics}\ }\textbf {\bibinfo {volume} {63}},\ \bibinfo {pages} {021903} (\bibinfo {year} {2001})}\BibitemShut {NoStop}%
\bibitem [{\citenamefont {Phogat}\ \emph {et~al.}(2025)\citenamefont {Phogat}, \citenamefont {Behler}, \citenamefont {Sonkusare}, \citenamefont {Pang}, \citenamefont {Koussis}, \citenamefont {Roberts}, \citenamefont {DeKraker}, \citenamefont {Shine}, \citenamefont {Fornito}, \citenamefont {Robinson},\ and\ \citenamefont {Breakspear}}]{phogat_unified_2025}%
  \BibitemOpen
  \bibfield  {author} {\bibinfo {author} {\bibfnamefont {R.}~\bibnamefont {Phogat}}, \bibinfo {author} {\bibfnamefont {A.}~\bibnamefont {Behler}}, \bibinfo {author} {\bibfnamefont {S.}~\bibnamefont {Sonkusare}}, \bibinfo {author} {\bibfnamefont {J.~C.}\ \bibnamefont {Pang}}, \bibinfo {author} {\bibfnamefont {N.}~\bibnamefont {Koussis}}, \bibinfo {author} {\bibfnamefont {J.~A.}\ \bibnamefont {Roberts}}, \bibinfo {author} {\bibfnamefont {J.}~\bibnamefont {DeKraker}}, \bibinfo {author} {\bibfnamefont {J.~M.}\ \bibnamefont {Shine}}, \bibinfo {author} {\bibfnamefont {A.}~\bibnamefont {Fornito}}, \bibinfo {author} {\bibfnamefont {P.~A.}\ \bibnamefont {Robinson}},\ and\ \bibinfo {author} {\bibfnamefont {M.}~\bibnamefont {Breakspear}},\ }\href {https://doi.org/10.1101/2025.09.07.674721} {\bibinfo {title} {A unified model of cortico-hippocampal interactions through neural field theory}} (\bibinfo {year} {2025})\BibitemShut {NoStop}%
\bibitem [{\citenamefont {Robinson}\ \emph {et~al.}(1997)\citenamefont {Robinson}, \citenamefont {Rennie},\ and\ \citenamefont {Wright}}]{robinson_propagation_1997}%
  \BibitemOpen
  \bibfield  {author} {\bibinfo {author} {\bibfnamefont {P.~A.}\ \bibnamefont {Robinson}}, \bibinfo {author} {\bibfnamefont {C.~J.}\ \bibnamefont {Rennie}},\ and\ \bibinfo {author} {\bibfnamefont {J.~J.}\ \bibnamefont {Wright}},\ }\href {https://doi.org/10.1103/PhysRevE.56.826} {\bibfield  {journal} {\bibinfo  {journal} {Physical Review E}\ }\textbf {\bibinfo {volume} {56}},\ \bibinfo {pages} {826} (\bibinfo {year} {1997})}\BibitemShut {NoStop}%
\bibitem [{\citenamefont {Fischl}(2012)}]{fischl_freesurfer_2012}%
  \BibitemOpen
  \bibfield  {author} {\bibinfo {author} {\bibfnamefont {B.}~\bibnamefont {Fischl}},\ }\href {https://doi.org/10.1016/j.neuroimage.2012.01.021} {\bibfield  {journal} {\bibinfo  {journal} {NeuroImage}\ }\textbf {\bibinfo {volume} {62}},\ \bibinfo {pages} {774} (\bibinfo {year} {2012})}\BibitemShut {NoStop}%
\bibitem [{\citenamefont {Sekihara}\ and\ \citenamefont {Nagarajan}(2015)}]{sekihara_electromagnetic_2015}%
  \BibitemOpen
  \bibfield  {author} {\bibinfo {author} {\bibfnamefont {K.}~\bibnamefont {Sekihara}}\ and\ \bibinfo {author} {\bibfnamefont {S.~S.}\ \bibnamefont {Nagarajan}},\ }\href {https://doi.org/10.1007/978-3-319-14947-9} {\emph {\bibinfo {title} {Electromagnetic {Brain} {Imaging}: {A} {Bayesian} {Perspective}}}}\ (\bibinfo  {publisher} {Springer International Publishing},\ \bibinfo {address} {Cham},\ \bibinfo {year} {2015})\BibitemShut {NoStop}%
\bibitem [{\citenamefont {Vincent}\ \emph {et~al.}(2007)\citenamefont {Vincent}, \citenamefont {Gribonval},\ and\ \citenamefont {Plumbley}}]{vincent_oracle_2007}%
  \BibitemOpen
  \bibfield  {author} {\bibinfo {author} {\bibfnamefont {E.}~\bibnamefont {Vincent}}, \bibinfo {author} {\bibfnamefont {R.}~\bibnamefont {Gribonval}},\ and\ \bibinfo {author} {\bibfnamefont {M.~D.}\ \bibnamefont {Plumbley}},\ }\href {https://doi.org/10.1016/j.sigpro.2007.01.016} {\bibfield  {journal} {\bibinfo  {journal} {Signal Processing}\ }\bibinfo {series} {Independent {Component} {Analysis} and {Blind} {Source} {Separation}},\ \textbf {\bibinfo {volume} {87}},\ \bibinfo {pages} {1933} (\bibinfo {year} {2007})}\BibitemShut {NoStop}%
\bibitem [{\citenamefont {Siu}\ \emph {et~al.}(2026{\natexlab{b}})\citenamefont {Siu}, \citenamefont {Karoly}, \citenamefont {Mansour~Lakouraj}, \citenamefont {Soto-Breceda}, \citenamefont {Kuhlmann}, \citenamefont {Cook},\ and\ \citenamefont {Grayden}}]{siu_neural_2026}%
  \BibitemOpen
  \bibfield  {author} {\bibinfo {author} {\bibfnamefont {P.~H.}\ \bibnamefont {Siu}}, \bibinfo {author} {\bibfnamefont {P.~J.}\ \bibnamefont {Karoly}}, \bibinfo {author} {\bibfnamefont {S.}~\bibnamefont {Mansour~Lakouraj}}, \bibinfo {author} {\bibfnamefont {A.}~\bibnamefont {Soto-Breceda}}, \bibinfo {author} {\bibfnamefont {L.}~\bibnamefont {Kuhlmann}}, \bibinfo {author} {\bibfnamefont {M.~J.}\ \bibnamefont {Cook}},\ and\ \bibinfo {author} {\bibfnamefont {D.~B.}\ \bibnamefont {Grayden}},\ }\bibfield  {journal} {\bibinfo  {journal} {Journal of Neural Engineering}\ }\textbf {\bibinfo {volume} {23}},\ \href {https://doi.org/10.1088/1741-2552/ae80fb} {10.1088/1741-2552/ae80fb} (\bibinfo {year} {2026}{\natexlab{b}})\BibitemShut {NoStop}%
\bibitem [{\citenamefont {He}\ \emph {et~al.}(1987)\citenamefont {He}, \citenamefont {Musha}, \citenamefont {Okamoto}, \citenamefont {Homma}, \citenamefont {Nakajima},\ and\ \citenamefont {Sato}}]{he_electric_1987}%
  \BibitemOpen
  \bibfield  {author} {\bibinfo {author} {\bibfnamefont {B.}~\bibnamefont {He}}, \bibinfo {author} {\bibfnamefont {T.}~\bibnamefont {Musha}}, \bibinfo {author} {\bibfnamefont {Y.}~\bibnamefont {Okamoto}}, \bibinfo {author} {\bibfnamefont {S.}~\bibnamefont {Homma}}, \bibinfo {author} {\bibfnamefont {Y.}~\bibnamefont {Nakajima}},\ and\ \bibinfo {author} {\bibfnamefont {T.}~\bibnamefont {Sato}},\ }\href {https://doi.org/10.1109/TBME.1987.326056} {\bibfield  {journal} {\bibinfo  {journal} {IEEE Transactions on Biomedical Engineering}\ }\textbf {\bibinfo {volume} {BME-34}},\ \bibinfo {pages} {406} (\bibinfo {year} {1987})}\BibitemShut {NoStop}%
\bibitem [{\citenamefont {Gramfort}\ \emph {et~al.}(2010)\citenamefont {Gramfort}, \citenamefont {Papadopoulo}, \citenamefont {Olivi},\ and\ \citenamefont {Clerc}}]{gramfort_openmeeg_2010}%
  \BibitemOpen
  \bibfield  {author} {\bibinfo {author} {\bibfnamefont {A.}~\bibnamefont {Gramfort}}, \bibinfo {author} {\bibfnamefont {T.}~\bibnamefont {Papadopoulo}}, \bibinfo {author} {\bibfnamefont {E.}~\bibnamefont {Olivi}},\ and\ \bibinfo {author} {\bibfnamefont {M.}~\bibnamefont {Clerc}},\ }\href {https://doi.org/10.1186/1475-925X-9-45} {\bibfield  {journal} {\bibinfo  {journal} {BioMedical Engineering OnLine}\ }\textbf {\bibinfo {volume} {9}},\ \bibinfo {pages} {45} (\bibinfo {year} {2010})}\BibitemShut {NoStop}%
\bibitem [{\citenamefont {Gramfort}\ \emph {et~al.}(2014)\citenamefont {Gramfort}, \citenamefont {Luessi}, \citenamefont {Larson}, \citenamefont {Engemann}, \citenamefont {Strohmeier}, \citenamefont {Brodbeck}, \citenamefont {Parkkonen},\ and\ \citenamefont {Hämäläinen}}]{gramfort_mne_2014}%
  \BibitemOpen
  \bibfield  {author} {\bibinfo {author} {\bibfnamefont {A.}~\bibnamefont {Gramfort}}, \bibinfo {author} {\bibfnamefont {M.}~\bibnamefont {Luessi}}, \bibinfo {author} {\bibfnamefont {E.}~\bibnamefont {Larson}}, \bibinfo {author} {\bibfnamefont {D.~A.}\ \bibnamefont {Engemann}}, \bibinfo {author} {\bibfnamefont {D.}~\bibnamefont {Strohmeier}}, \bibinfo {author} {\bibfnamefont {C.}~\bibnamefont {Brodbeck}}, \bibinfo {author} {\bibfnamefont {L.}~\bibnamefont {Parkkonen}},\ and\ \bibinfo {author} {\bibfnamefont {M.~S.}\ \bibnamefont {Hämäläinen}},\ }\href {https://doi.org/10.1016/j.neuroimage.2013.10.027} {\bibfield  {journal} {\bibinfo  {journal} {NeuroImage}\ }\textbf {\bibinfo {volume} {86}},\ \bibinfo {pages} {446} (\bibinfo {year} {2014})}\BibitemShut {NoStop}%
\bibitem [{\citenamefont {Mikulan}\ \emph {et~al.}(2020)\citenamefont {Mikulan}, \citenamefont {Russo}, \citenamefont {Parmigiani}, \citenamefont {Sarasso}, \citenamefont {Zauli}, \citenamefont {Rubino}, \citenamefont {Avanzini}, \citenamefont {Cattani}, \citenamefont {Sorrentino}, \citenamefont {Gibbs}, \citenamefont {Cardinale}, \citenamefont {Sartori}, \citenamefont {Nobili}, \citenamefont {Massimini},\ and\ \citenamefont {Pigorini}}]{mikulan_simultaneous_2020}%
  \BibitemOpen
  \bibfield  {author} {\bibinfo {author} {\bibfnamefont {E.}~\bibnamefont {Mikulan}}, \bibinfo {author} {\bibfnamefont {S.}~\bibnamefont {Russo}}, \bibinfo {author} {\bibfnamefont {S.}~\bibnamefont {Parmigiani}}, \bibinfo {author} {\bibfnamefont {S.}~\bibnamefont {Sarasso}}, \bibinfo {author} {\bibfnamefont {F.~M.}\ \bibnamefont {Zauli}}, \bibinfo {author} {\bibfnamefont {A.}~\bibnamefont {Rubino}}, \bibinfo {author} {\bibfnamefont {P.}~\bibnamefont {Avanzini}}, \bibinfo {author} {\bibfnamefont {A.}~\bibnamefont {Cattani}}, \bibinfo {author} {\bibfnamefont {A.}~\bibnamefont {Sorrentino}}, \bibinfo {author} {\bibfnamefont {S.}~\bibnamefont {Gibbs}}, \bibinfo {author} {\bibfnamefont {F.}~\bibnamefont {Cardinale}}, \bibinfo {author} {\bibfnamefont {I.}~\bibnamefont {Sartori}}, \bibinfo {author} {\bibfnamefont {L.}~\bibnamefont {Nobili}}, \bibinfo {author} {\bibfnamefont {M.}~\bibnamefont {Massimini}},\ and\ \bibinfo {author} {\bibfnamefont {A.}~\bibnamefont {Pigorini}},\ }\href
  {https://doi.org/10.1038/s41597-020-0467-x} {\bibfield  {journal} {\bibinfo  {journal} {Scientific Data}\ }\textbf {\bibinfo {volume} {7}},\ \bibinfo {pages} {127} (\bibinfo {year} {2020})},\ \bibinfo {note} {number: 1}\BibitemShut {NoStop}%
\bibitem [{\citenamefont {Parmigiani}\ \emph {et~al.}(2022)\citenamefont {Parmigiani}, \citenamefont {Mikulan}, \citenamefont {Russo}, \citenamefont {Sarasso}, \citenamefont {Zauli}, \citenamefont {Rubino}, \citenamefont {Cattani}, \citenamefont {Fecchio}, \citenamefont {Giampiccolo}, \citenamefont {Lanzone}, \citenamefont {D'Orio}, \citenamefont {Del~Vecchio}, \citenamefont {Avanzini}, \citenamefont {Nobili}, \citenamefont {Sartori}, \citenamefont {Massimini},\ and\ \citenamefont {Pigorini}}]{parmigiani_simultaneous_2022}%
  \BibitemOpen
  \bibfield  {author} {\bibinfo {author} {\bibfnamefont {S.}~\bibnamefont {Parmigiani}}, \bibinfo {author} {\bibfnamefont {E.}~\bibnamefont {Mikulan}}, \bibinfo {author} {\bibfnamefont {S.}~\bibnamefont {Russo}}, \bibinfo {author} {\bibfnamefont {S.}~\bibnamefont {Sarasso}}, \bibinfo {author} {\bibfnamefont {F.~M.}\ \bibnamefont {Zauli}}, \bibinfo {author} {\bibfnamefont {A.}~\bibnamefont {Rubino}}, \bibinfo {author} {\bibfnamefont {A.}~\bibnamefont {Cattani}}, \bibinfo {author} {\bibfnamefont {M.}~\bibnamefont {Fecchio}}, \bibinfo {author} {\bibfnamefont {D.}~\bibnamefont {Giampiccolo}}, \bibinfo {author} {\bibfnamefont {J.}~\bibnamefont {Lanzone}}, \bibinfo {author} {\bibfnamefont {P.}~\bibnamefont {D'Orio}}, \bibinfo {author} {\bibfnamefont {M.}~\bibnamefont {Del~Vecchio}}, \bibinfo {author} {\bibfnamefont {P.}~\bibnamefont {Avanzini}}, \bibinfo {author} {\bibfnamefont {L.}~\bibnamefont {Nobili}}, \bibinfo {author} {\bibfnamefont {I.}~\bibnamefont {Sartori}}, \bibinfo {author} {\bibfnamefont
  {M.}~\bibnamefont {Massimini}},\ and\ \bibinfo {author} {\bibfnamefont {A.}~\bibnamefont {Pigorini}},\ }\href {https://doi.org/10.1016/j.brs.2022.04.007} {\bibfield  {journal} {\bibinfo  {journal} {Brain Stimulation}\ }\textbf {\bibinfo {volume} {15}},\ \bibinfo {pages} {664} (\bibinfo {year} {2022})}\BibitemShut {NoStop}%
\bibitem [{\citenamefont {Li}\ \emph {et~al.}(2025)\citenamefont {Li}, \citenamefont {Zalesky}, \citenamefont {Wang}, \citenamefont {Wang}, \citenamefont {Ma}, \citenamefont {Cheng}, \citenamefont {Banaschewski}, \citenamefont {Barker}, \citenamefont {Bokde}, \citenamefont {Brühl}, \citenamefont {Desrivières}, \citenamefont {Flor}, \citenamefont {Garavan}, \citenamefont {Gowland}, \citenamefont {Grigis}, \citenamefont {Heinz}, \citenamefont {Lemaître}, \citenamefont {Martinot}, \citenamefont {Martinot}, \citenamefont {Artiges}, \citenamefont {Nees}, \citenamefont {Orfanos}, \citenamefont {Poustka}, \citenamefont {Smolka}, \citenamefont {Vaidya}, \citenamefont {Walter}, \citenamefont {Whelan}, \citenamefont {Schumann}, \citenamefont {Jia}, \citenamefont {Chu},\ and\ \citenamefont {Fan}}]{li_mapping_2025}%
  \BibitemOpen
  \bibfield  {author} {\bibinfo {author} {\bibfnamefont {D.}~\bibnamefont {Li}}, \bibinfo {author} {\bibfnamefont {A.}~\bibnamefont {Zalesky}}, \bibinfo {author} {\bibfnamefont {Y.}~\bibnamefont {Wang}}, \bibinfo {author} {\bibfnamefont {H.}~\bibnamefont {Wang}}, \bibinfo {author} {\bibfnamefont {L.}~\bibnamefont {Ma}}, \bibinfo {author} {\bibfnamefont {L.}~\bibnamefont {Cheng}}, \bibinfo {author} {\bibfnamefont {T.}~\bibnamefont {Banaschewski}}, \bibinfo {author} {\bibfnamefont {G.~J.}\ \bibnamefont {Barker}}, \bibinfo {author} {\bibfnamefont {A.~L.~W.}\ \bibnamefont {Bokde}}, \bibinfo {author} {\bibfnamefont {R.}~\bibnamefont {Brühl}}, \bibinfo {author} {\bibfnamefont {S.}~\bibnamefont {Desrivières}}, \bibinfo {author} {\bibfnamefont {H.}~\bibnamefont {Flor}}, \bibinfo {author} {\bibfnamefont {H.}~\bibnamefont {Garavan}}, \bibinfo {author} {\bibfnamefont {P.}~\bibnamefont {Gowland}}, \bibinfo {author} {\bibfnamefont {A.}~\bibnamefont {Grigis}}, \bibinfo {author} {\bibfnamefont {A.}~\bibnamefont {Heinz}},
  \bibinfo {author} {\bibfnamefont {H.}~\bibnamefont {Lemaître}}, \bibinfo {author} {\bibfnamefont {J.-L.}\ \bibnamefont {Martinot}}, \bibinfo {author} {\bibfnamefont {M.-L.~P.}\ \bibnamefont {Martinot}}, \bibinfo {author} {\bibfnamefont {E.}~\bibnamefont {Artiges}}, \bibinfo {author} {\bibfnamefont {F.}~\bibnamefont {Nees}}, \bibinfo {author} {\bibfnamefont {D.~P.}\ \bibnamefont {Orfanos}}, \bibinfo {author} {\bibfnamefont {L.}~\bibnamefont {Poustka}}, \bibinfo {author} {\bibfnamefont {M.~N.}\ \bibnamefont {Smolka}}, \bibinfo {author} {\bibfnamefont {N.}~\bibnamefont {Vaidya}}, \bibinfo {author} {\bibfnamefont {H.}~\bibnamefont {Walter}}, \bibinfo {author} {\bibfnamefont {R.}~\bibnamefont {Whelan}}, \bibinfo {author} {\bibfnamefont {G.}~\bibnamefont {Schumann}}, \bibinfo {author} {\bibfnamefont {T.}~\bibnamefont {Jia}}, \bibinfo {author} {\bibfnamefont {C.}~\bibnamefont {Chu}},\ and\ \bibinfo {author} {\bibfnamefont {L.}~\bibnamefont {Fan}},\ }\href {https://doi.org/10.1038/s41467-025-62812-9} {\bibfield
  {journal} {\bibinfo  {journal} {Nature Communications}\ }\textbf {\bibinfo {volume} {16}},\ \bibinfo {pages} {7489} (\bibinfo {year} {2025})}\BibitemShut {NoStop}%
\bibitem [{\citenamefont {Potash}\ \emph {et~al.}(2025)\citenamefont {Potash}, \citenamefont {Yang}, \citenamefont {Winston}, \citenamefont {Atasoy}, \citenamefont {Kringelbach}, \citenamefont {Sparby},\ and\ \citenamefont {Sacchet}}]{potash_investigating_2025}%
  \BibitemOpen
  \bibfield  {author} {\bibinfo {author} {\bibfnamefont {R.~M.}\ \bibnamefont {Potash}}, \bibinfo {author} {\bibfnamefont {W.~F.~Z.}\ \bibnamefont {Yang}}, \bibinfo {author} {\bibfnamefont {B.}~\bibnamefont {Winston}}, \bibinfo {author} {\bibfnamefont {S.}~\bibnamefont {Atasoy}}, \bibinfo {author} {\bibfnamefont {M.~L.}\ \bibnamefont {Kringelbach}}, \bibinfo {author} {\bibfnamefont {T.}~\bibnamefont {Sparby}},\ and\ \bibinfo {author} {\bibfnamefont {M.~D.}\ \bibnamefont {Sacchet}},\ }\href {https://doi.org/10.1093/cercor/bhaf039} {\bibfield  {journal} {\bibinfo  {journal} {Cerebral Cortex}\ }\textbf {\bibinfo {volume} {35}},\ \bibinfo {pages} {bhaf039} (\bibinfo {year} {2025})}\BibitemShut {NoStop}%
\bibitem [{\citenamefont {Glover}(2011)}]{glover_overview_2011}%
  \BibitemOpen
  \bibfield  {author} {\bibinfo {author} {\bibfnamefont {G.~H.}\ \bibnamefont {Glover}},\ }\href {https://doi.org/10.1016/j.nec.2010.11.001} {\bibfield  {journal} {\bibinfo  {journal} {Neurosurgery Clinics of North America}\ }\bibinfo {series} {Functional {Imaging}},\ \textbf {\bibinfo {volume} {22}},\ \bibinfo {pages} {133} (\bibinfo {year} {2011})}\BibitemShut {NoStop}%
\bibitem [{\citenamefont {Van~Essen}\ \emph {et~al.}(2013)\citenamefont {Van~Essen}, \citenamefont {Smith}, \citenamefont {Barch}, \citenamefont {Behrens}, \citenamefont {Yacoub},\ and\ \citenamefont {Ugurbil}}]{van_essen_wu-minn_2013}%
  \BibitemOpen
  \bibfield  {author} {\bibinfo {author} {\bibfnamefont {D.~C.}\ \bibnamefont {Van~Essen}}, \bibinfo {author} {\bibfnamefont {S.~M.}\ \bibnamefont {Smith}}, \bibinfo {author} {\bibfnamefont {D.~M.}\ \bibnamefont {Barch}}, \bibinfo {author} {\bibfnamefont {T.~E.~J.}\ \bibnamefont {Behrens}}, \bibinfo {author} {\bibfnamefont {E.}~\bibnamefont {Yacoub}},\ and\ \bibinfo {author} {\bibfnamefont {K.}~\bibnamefont {Ugurbil}},\ }\href {https://doi.org/10.1016/j.neuroimage.2013.05.041} {\bibfield  {journal} {\bibinfo  {journal} {NeuroImage}\ }\bibinfo {series} {Mapping the {Connectome}},\ \textbf {\bibinfo {volume} {80}},\ \bibinfo {pages} {62} (\bibinfo {year} {2013})}\BibitemShut {NoStop}%
\bibitem [{\citenamefont {Sun}\ \emph {et~al.}(2025)\citenamefont {Sun}, \citenamefont {Jing}, \citenamefont {Turley}, \citenamefont {Alcott}, \citenamefont {Kang}, \citenamefont {Cole}, \citenamefont {Goldenholz}, \citenamefont {Lam}, \citenamefont {Amorim}, \citenamefont {Chu}, \citenamefont {Cash}, \citenamefont {Junior}, \citenamefont {Gupta}, \citenamefont {Ghanta}, \citenamefont {Nearing}, \citenamefont {Nascimento}, \citenamefont {Struck}, \citenamefont {Kim}, \citenamefont {Sartipi}, \citenamefont {Tauton}, \citenamefont {Fernandes}, \citenamefont {Sun}, \citenamefont {Bayas}, \citenamefont {Gallagher}, \citenamefont {Wagenaar}, \citenamefont {Sinha}, \citenamefont {Lee-Messer}, \citenamefont {Silvers}, \citenamefont {Gunapati}, \citenamefont {Rosand}, \citenamefont {Peters}, \citenamefont {Loddenkemper}, \citenamefont {Lee}, \citenamefont {Zafar},\ and\ \citenamefont {Westover}}]{sun_harvard_2025}%
  \BibitemOpen
  \bibfield  {author} {\bibinfo {author} {\bibfnamefont {C.}~\bibnamefont {Sun}}, \bibinfo {author} {\bibfnamefont {J.}~\bibnamefont {Jing}}, \bibinfo {author} {\bibfnamefont {N.}~\bibnamefont {Turley}}, \bibinfo {author} {\bibfnamefont {C.}~\bibnamefont {Alcott}}, \bibinfo {author} {\bibfnamefont {W.-Y.}\ \bibnamefont {Kang}}, \bibinfo {author} {\bibfnamefont {A.~J.}\ \bibnamefont {Cole}}, \bibinfo {author} {\bibfnamefont {D.~M.}\ \bibnamefont {Goldenholz}}, \bibinfo {author} {\bibfnamefont {A.}~\bibnamefont {Lam}}, \bibinfo {author} {\bibfnamefont {E.}~\bibnamefont {Amorim}}, \bibinfo {author} {\bibfnamefont {C.}~\bibnamefont {Chu}}, \bibinfo {author} {\bibfnamefont {S.}~\bibnamefont {Cash}}, \bibinfo {author} {\bibfnamefont {V.~M.}\ \bibnamefont {Junior}}, \bibinfo {author} {\bibfnamefont {A.}~\bibnamefont {Gupta}}, \bibinfo {author} {\bibfnamefont {M.}~\bibnamefont {Ghanta}}, \bibinfo {author} {\bibfnamefont {B.}~\bibnamefont {Nearing}}, \bibinfo {author} {\bibfnamefont {F.~A.}\ \bibnamefont
  {Nascimento}}, \bibinfo {author} {\bibfnamefont {A.}~\bibnamefont {Struck}}, \bibinfo {author} {\bibfnamefont {J.}~\bibnamefont {Kim}}, \bibinfo {author} {\bibfnamefont {S.}~\bibnamefont {Sartipi}}, \bibinfo {author} {\bibfnamefont {A.-M.}\ \bibnamefont {Tauton}}, \bibinfo {author} {\bibfnamefont {M.}~\bibnamefont {Fernandes}}, \bibinfo {author} {\bibfnamefont {H.}~\bibnamefont {Sun}}, \bibinfo {author} {\bibfnamefont {G.}~\bibnamefont {Bayas}}, \bibinfo {author} {\bibfnamefont {K.}~\bibnamefont {Gallagher}}, \bibinfo {author} {\bibfnamefont {J.~B.}\ \bibnamefont {Wagenaar}}, \bibinfo {author} {\bibfnamefont {N.}~\bibnamefont {Sinha}}, \bibinfo {author} {\bibfnamefont {C.}~\bibnamefont {Lee-Messer}}, \bibinfo {author} {\bibfnamefont {C.~T.}\ \bibnamefont {Silvers}}, \bibinfo {author} {\bibfnamefont {B.}~\bibnamefont {Gunapati}}, \bibinfo {author} {\bibfnamefont {J.}~\bibnamefont {Rosand}}, \bibinfo {author} {\bibfnamefont {J.}~\bibnamefont {Peters}}, \bibinfo {author} {\bibfnamefont {T.}~\bibnamefont
  {Loddenkemper}}, \bibinfo {author} {\bibfnamefont {J.~W.}\ \bibnamefont {Lee}}, \bibinfo {author} {\bibfnamefont {S.}~\bibnamefont {Zafar}},\ and\ \bibinfo {author} {\bibfnamefont {M.~B.}\ \bibnamefont {Westover}},\ }\bibfield  {journal} {\bibinfo  {journal} {Epilepsia}\ }\textbf {\bibinfo {volume} {66}},\ \href {https://doi.org/10.1111/epi.18487} {10.1111/epi.18487} (\bibinfo {year} {2025}),\ \bibinfo {note} {\_eprint: https://onlinelibrary.wiley.com/doi/pdf/10.1111/epi.18487}\BibitemShut {NoStop}%
\bibitem [{\citenamefont {Jirsa}\ \emph {et~al.}(2014)\citenamefont {Jirsa}, \citenamefont {Stacey}, \citenamefont {Quilichini}, \citenamefont {Ivanov},\ and\ \citenamefont {Bernard}}]{jirsa_nature_2014}%
  \BibitemOpen
  \bibfield  {author} {\bibinfo {author} {\bibfnamefont {V.~K.}\ \bibnamefont {Jirsa}}, \bibinfo {author} {\bibfnamefont {W.~C.}\ \bibnamefont {Stacey}}, \bibinfo {author} {\bibfnamefont {P.~P.}\ \bibnamefont {Quilichini}}, \bibinfo {author} {\bibfnamefont {A.~I.}\ \bibnamefont {Ivanov}},\ and\ \bibinfo {author} {\bibfnamefont {C.}~\bibnamefont {Bernard}},\ }\href {https://doi.org/10.1093/brain/awu133} {\bibfield  {journal} {\bibinfo  {journal} {Brain}\ }\textbf {\bibinfo {volume} {137}},\ \bibinfo {pages} {2210} (\bibinfo {year} {2014})}\BibitemShut {NoStop}%
\bibitem [{\citenamefont {Saggio}\ \emph {et~al.}(2020)\citenamefont {Saggio}, \citenamefont {Crisp}, \citenamefont {Scott}, \citenamefont {Karoly}, \citenamefont {Kuhlmann}, \citenamefont {Nakatani}, \citenamefont {Murai}, \citenamefont {Dümpelmann}, \citenamefont {Schulze-Bonhage}, \citenamefont {Ikeda}, \citenamefont {Cook}, \citenamefont {Gliske}, \citenamefont {Lin}, \citenamefont {Bernard}, \citenamefont {Jirsa},\ and\ \citenamefont {Stacey}}]{saggio_taxonomy_2020}%
  \BibitemOpen
  \bibfield  {author} {\bibinfo {author} {\bibfnamefont {M.~L.}\ \bibnamefont {Saggio}}, \bibinfo {author} {\bibfnamefont {D.}~\bibnamefont {Crisp}}, \bibinfo {author} {\bibfnamefont {J.~M.}\ \bibnamefont {Scott}}, \bibinfo {author} {\bibfnamefont {P.}~\bibnamefont {Karoly}}, \bibinfo {author} {\bibfnamefont {L.}~\bibnamefont {Kuhlmann}}, \bibinfo {author} {\bibfnamefont {M.}~\bibnamefont {Nakatani}}, \bibinfo {author} {\bibfnamefont {T.}~\bibnamefont {Murai}}, \bibinfo {author} {\bibfnamefont {M.}~\bibnamefont {Dümpelmann}}, \bibinfo {author} {\bibfnamefont {A.}~\bibnamefont {Schulze-Bonhage}}, \bibinfo {author} {\bibfnamefont {A.}~\bibnamefont {Ikeda}}, \bibinfo {author} {\bibfnamefont {M.}~\bibnamefont {Cook}}, \bibinfo {author} {\bibfnamefont {S.~V.}\ \bibnamefont {Gliske}}, \bibinfo {author} {\bibfnamefont {J.}~\bibnamefont {Lin}}, \bibinfo {author} {\bibfnamefont {C.}~\bibnamefont {Bernard}}, \bibinfo {author} {\bibfnamefont {V.}~\bibnamefont {Jirsa}},\ and\ \bibinfo {author} {\bibfnamefont {W.~C.}\
  \bibnamefont {Stacey}},\ }\href {https://doi.org/10.7554/eLife.55632} {\bibfield  {journal} {\bibinfo  {journal} {eLife}\ }\textbf {\bibinfo {volume} {9}},\ \bibinfo {pages} {e55632} (\bibinfo {year} {2020})}\BibitemShut {NoStop}%
\bibitem [{\citenamefont {Halliday}\ \emph {et~al.}(2025)\citenamefont {Halliday}, \citenamefont {Gillinder}, \citenamefont {Lai}, \citenamefont {Seneviratne}, \citenamefont {Fontenot}, \citenamefont {Cameron}, \citenamefont {McLean}, \citenamefont {Niemiec}, \citenamefont {Raghupathi}, \citenamefont {Ganguly}, \citenamefont {Ellis}, \citenamefont {Conrad}, \citenamefont {Briggs}, \citenamefont {Bulluss}, \citenamefont {Kwan}, \citenamefont {Perucca}, \citenamefont {O'Brien}, \citenamefont {McGonigal}, \citenamefont {Gutman}, \citenamefont {Papacostas}, \citenamefont {Fong}, \citenamefont {Lee}, \citenamefont {Crompton}, \citenamefont {Laing}, \citenamefont {Wijayath}, \citenamefont {Morokoff}, \citenamefont {Murphy}, \citenamefont {D'Souza},\ and\ \citenamefont {Cook}}]{halliday_umpire_2025}%
  \BibitemOpen
  \bibfield  {author} {\bibinfo {author} {\bibfnamefont {A.~J.}\ \bibnamefont {Halliday}}, \bibinfo {author} {\bibfnamefont {L.}~\bibnamefont {Gillinder}}, \bibinfo {author} {\bibfnamefont {A.}~\bibnamefont {Lai}}, \bibinfo {author} {\bibfnamefont {U.}~\bibnamefont {Seneviratne}}, \bibinfo {author} {\bibfnamefont {H.}~\bibnamefont {Fontenot}}, \bibinfo {author} {\bibfnamefont {T.}~\bibnamefont {Cameron}}, \bibinfo {author} {\bibfnamefont {K.}~\bibnamefont {McLean}}, \bibinfo {author} {\bibfnamefont {A.}~\bibnamefont {Niemiec}}, \bibinfo {author} {\bibfnamefont {R.}~\bibnamefont {Raghupathi}}, \bibinfo {author} {\bibfnamefont {T.~M.}\ \bibnamefont {Ganguly}}, \bibinfo {author} {\bibfnamefont {C.}~\bibnamefont {Ellis}}, \bibinfo {author} {\bibfnamefont {E.~C.}\ \bibnamefont {Conrad}}, \bibinfo {author} {\bibfnamefont {R.}~\bibnamefont {Briggs}}, \bibinfo {author} {\bibfnamefont {K.}~\bibnamefont {Bulluss}}, \bibinfo {author} {\bibfnamefont {P.}~\bibnamefont {Kwan}}, \bibinfo {author} {\bibfnamefont
  {P.}~\bibnamefont {Perucca}}, \bibinfo {author} {\bibfnamefont {T.~J.}\ \bibnamefont {O'Brien}}, \bibinfo {author} {\bibfnamefont {A.}~\bibnamefont {McGonigal}}, \bibinfo {author} {\bibfnamefont {M.}~\bibnamefont {Gutman}}, \bibinfo {author} {\bibfnamefont {J.}~\bibnamefont {Papacostas}}, \bibinfo {author} {\bibfnamefont {M.~W.~K.}\ \bibnamefont {Fong}}, \bibinfo {author} {\bibfnamefont {A.}~\bibnamefont {Lee}}, \bibinfo {author} {\bibfnamefont {D.~E.}\ \bibnamefont {Crompton}}, \bibinfo {author} {\bibfnamefont {J.}~\bibnamefont {Laing}}, \bibinfo {author} {\bibfnamefont {M.}~\bibnamefont {Wijayath}}, \bibinfo {author} {\bibfnamefont {A.~P.}\ \bibnamefont {Morokoff}}, \bibinfo {author} {\bibfnamefont {M.}~\bibnamefont {Murphy}}, \bibinfo {author} {\bibfnamefont {W.~J.}\ \bibnamefont {D'Souza}},\ and\ \bibinfo {author} {\bibfnamefont {M.~J.}\ \bibnamefont {Cook}},\ }\href {https://doi.org/10.1111/epi.18458} {\bibfield  {journal} {\bibinfo  {journal} {Epilepsia}\ }\textbf {\bibinfo {volume} {66}},\ \bibinfo
  {pages} {3426} (\bibinfo {year} {2025})},\ \bibinfo {note} {\_eprint: https://onlinelibrary.wiley.com/doi/pdf/10.1111/epi.18458}\BibitemShut {NoStop}%
\bibitem [{\citenamefont {Sanger}\ \emph {et~al.}(2026)\citenamefont {Sanger}, \citenamefont {Zhang}, \citenamefont {Lisko}, \citenamefont {Farooqi}, \citenamefont {Ventz}, \citenamefont {Pati}, \citenamefont {Henry}, \citenamefont {Darrow}, \citenamefont {Park}, \citenamefont {Netoff},\ and\ \citenamefont {McGovern}}]{sanger_gamma_2026}%
  \BibitemOpen
  \bibfield  {author} {\bibinfo {author} {\bibfnamefont {Z.~T.}\ \bibnamefont {Sanger}}, \bibinfo {author} {\bibfnamefont {X.}~\bibnamefont {Zhang}}, \bibinfo {author} {\bibfnamefont {T.}~\bibnamefont {Lisko}}, \bibinfo {author} {\bibfnamefont {H.}~\bibnamefont {Farooqi}}, \bibinfo {author} {\bibfnamefont {S.}~\bibnamefont {Ventz}}, \bibinfo {author} {\bibfnamefont {S.}~\bibnamefont {Pati}}, \bibinfo {author} {\bibfnamefont {T.~R.}\ \bibnamefont {Henry}}, \bibinfo {author} {\bibfnamefont {D.~P.}\ \bibnamefont {Darrow}}, \bibinfo {author} {\bibfnamefont {M.~C.}\ \bibnamefont {Park}}, \bibinfo {author} {\bibfnamefont {T.~I.}\ \bibnamefont {Netoff}},\ and\ \bibinfo {author} {\bibfnamefont {R.~A.}\ \bibnamefont {McGovern}},\ }\bibfield  {journal} {\bibinfo  {journal} {Epilepsia}\ }\textbf {\bibinfo {volume} {n/a}},\ \href {https://doi.org/10.1002/epi.70340} {10.1002/epi.70340} (\bibinfo {year} {2026})\BibitemShut {NoStop}%
\bibitem [{\citenamefont {Lei}\ \emph {et~al.}(2015)\citenamefont {Lei}, \citenamefont {Wu},\ and\ \citenamefont {Valdes-Sosa}}]{lei_incorporating_2015}%
  \BibitemOpen
  \bibfield  {author} {\bibinfo {author} {\bibfnamefont {X.}~\bibnamefont {Lei}}, \bibinfo {author} {\bibfnamefont {T.}~\bibnamefont {Wu}},\ and\ \bibinfo {author} {\bibfnamefont {P.~A.}\ \bibnamefont {Valdes-Sosa}},\ }\bibfield  {journal} {\bibinfo  {journal} {Frontiers in Neuroscience}\ }\textbf {\bibinfo {volume} {9}},\ \href {https://doi.org/10.3389/fnins.2015.00284} {10.3389/fnins.2015.00284} (\bibinfo {year} {2015})\BibitemShut {NoStop}%
\bibitem [{\citenamefont {Hosseini}\ \emph {et~al.}(2018)\citenamefont {Hosseini}, \citenamefont {Sohrabpour},\ and\ \citenamefont {He}}]{hosseini_electromagnetic_2018}%
  \BibitemOpen
  \bibfield  {author} {\bibinfo {author} {\bibfnamefont {S.~A.~H.}\ \bibnamefont {Hosseini}}, \bibinfo {author} {\bibfnamefont {A.}~\bibnamefont {Sohrabpour}},\ and\ \bibinfo {author} {\bibfnamefont {B.}~\bibnamefont {He}},\ }\href {https://doi.org/10.1016/j.clinph.2017.10.027} {\bibfield  {journal} {\bibinfo  {journal} {Clinical Neurophysiology}\ }\textbf {\bibinfo {volume} {129}},\ \bibinfo {pages} {168} (\bibinfo {year} {2018})}\BibitemShut {NoStop}%
\bibitem [{\citenamefont {Molins}\ \emph {et~al.}(2008)\citenamefont {Molins}, \citenamefont {Stufflebeam}, \citenamefont {Brown},\ and\ \citenamefont {Hämäläinen}}]{molins_quantification_2008}%
  \BibitemOpen
  \bibfield  {author} {\bibinfo {author} {\bibfnamefont {A.}~\bibnamefont {Molins}}, \bibinfo {author} {\bibfnamefont {S.~M.}\ \bibnamefont {Stufflebeam}}, \bibinfo {author} {\bibfnamefont {E.~N.}\ \bibnamefont {Brown}},\ and\ \bibinfo {author} {\bibfnamefont {M.~S.}\ \bibnamefont {Hämäläinen}},\ }\href {https://doi.org/10.1016/j.neuroimage.2008.05.064} {\bibfield  {journal} {\bibinfo  {journal} {NeuroImage}\ }\textbf {\bibinfo {volume} {42}},\ \bibinfo {pages} {1069} (\bibinfo {year} {2008})}\BibitemShut {NoStop}%
\bibitem [{\citenamefont {He}\ and\ \citenamefont {Liu}(2008)}]{he_multimodal_2008}%
  \BibitemOpen
  \bibfield  {author} {\bibinfo {author} {\bibfnamefont {B.}~\bibnamefont {He}}\ and\ \bibinfo {author} {\bibfnamefont {Z.}~\bibnamefont {Liu}},\ }\href {https://doi.org/10.1109/RBME.2008.2008233} {\bibfield  {journal} {\bibinfo  {journal} {IEEE Reviews in Biomedical Engineering}\ }\textbf {\bibinfo {volume} {1}},\ \bibinfo {pages} {23} (\bibinfo {year} {2008})},\ \bibinfo {note} {conference Name: IEEE Reviews in Biomedical Engineering}\BibitemShut {NoStop}%
\bibitem [{\citenamefont {Yan}\ \emph {et~al.}(2023)\citenamefont {Yan}, \citenamefont {Kong}, \citenamefont {Xue}, \citenamefont {Yang}, \citenamefont {Orban}, \citenamefont {An}, \citenamefont {Holmes}, \citenamefont {Qian}, \citenamefont {Chen}, \citenamefont {Zuo}, \citenamefont {Zhou}, \citenamefont {Fortier}, \citenamefont {Tan}, \citenamefont {Gluckman}, \citenamefont {Chong}, \citenamefont {Meaney}, \citenamefont {Bzdok}, \citenamefont {Eickhoff},\ and\ \citenamefont {Yeo}}]{yan_homotopic_2023}%
  \BibitemOpen
  \bibfield  {author} {\bibinfo {author} {\bibfnamefont {X.}~\bibnamefont {Yan}}, \bibinfo {author} {\bibfnamefont {R.}~\bibnamefont {Kong}}, \bibinfo {author} {\bibfnamefont {A.}~\bibnamefont {Xue}}, \bibinfo {author} {\bibfnamefont {Q.}~\bibnamefont {Yang}}, \bibinfo {author} {\bibfnamefont {C.}~\bibnamefont {Orban}}, \bibinfo {author} {\bibfnamefont {L.}~\bibnamefont {An}}, \bibinfo {author} {\bibfnamefont {A.~J.}\ \bibnamefont {Holmes}}, \bibinfo {author} {\bibfnamefont {X.}~\bibnamefont {Qian}}, \bibinfo {author} {\bibfnamefont {J.}~\bibnamefont {Chen}}, \bibinfo {author} {\bibfnamefont {X.-N.}\ \bibnamefont {Zuo}}, \bibinfo {author} {\bibfnamefont {J.~H.}\ \bibnamefont {Zhou}}, \bibinfo {author} {\bibfnamefont {M.~V.}\ \bibnamefont {Fortier}}, \bibinfo {author} {\bibfnamefont {A.~P.}\ \bibnamefont {Tan}}, \bibinfo {author} {\bibfnamefont {P.}~\bibnamefont {Gluckman}}, \bibinfo {author} {\bibfnamefont {Y.~S.}\ \bibnamefont {Chong}}, \bibinfo {author} {\bibfnamefont {M.~J.}\ \bibnamefont {Meaney}},
  \bibinfo {author} {\bibfnamefont {D.}~\bibnamefont {Bzdok}}, \bibinfo {author} {\bibfnamefont {S.~B.}\ \bibnamefont {Eickhoff}},\ and\ \bibinfo {author} {\bibfnamefont {B.~T.}\ \bibnamefont {Yeo}},\ }\href {https://doi.org/10.1016/j.neuroimage.2023.120010} {\bibfield  {journal} {\bibinfo  {journal} {NeuroImage}\ }\textbf {\bibinfo {volume} {273}},\ \bibinfo {pages} {120010} (\bibinfo {year} {2023})}\BibitemShut {NoStop}%
\bibitem [{\citenamefont {Glasser}\ \emph {et~al.}(2013)\citenamefont {Glasser}, \citenamefont {Sotiropoulos}, \citenamefont {Wilson}, \citenamefont {Coalson}, \citenamefont {Fischl}, \citenamefont {Andersson}, \citenamefont {Xu}, \citenamefont {Jbabdi}, \citenamefont {Webster}, \citenamefont {Polimeni}, \citenamefont {Van~Essen},\ and\ \citenamefont {Jenkinson}}]{glasser_minimal_2013}%
  \BibitemOpen
  \bibfield  {author} {\bibinfo {author} {\bibfnamefont {M.~F.}\ \bibnamefont {Glasser}}, \bibinfo {author} {\bibfnamefont {S.~N.}\ \bibnamefont {Sotiropoulos}}, \bibinfo {author} {\bibfnamefont {J.~A.}\ \bibnamefont {Wilson}}, \bibinfo {author} {\bibfnamefont {T.~S.}\ \bibnamefont {Coalson}}, \bibinfo {author} {\bibfnamefont {B.}~\bibnamefont {Fischl}}, \bibinfo {author} {\bibfnamefont {J.~L.}\ \bibnamefont {Andersson}}, \bibinfo {author} {\bibfnamefont {J.}~\bibnamefont {Xu}}, \bibinfo {author} {\bibfnamefont {S.}~\bibnamefont {Jbabdi}}, \bibinfo {author} {\bibfnamefont {M.}~\bibnamefont {Webster}}, \bibinfo {author} {\bibfnamefont {J.~R.}\ \bibnamefont {Polimeni}}, \bibinfo {author} {\bibfnamefont {D.~C.}\ \bibnamefont {Van~Essen}},\ and\ \bibinfo {author} {\bibfnamefont {M.}~\bibnamefont {Jenkinson}},\ }\href {https://doi.org/10.1016/j.neuroimage.2013.04.127} {\bibfield  {journal} {\bibinfo  {journal} {NeuroImage}\ }\textbf {\bibinfo {volume} {80}},\ \bibinfo {pages} {105} (\bibinfo {year}
  {2013})}\BibitemShut {NoStop}%
\bibitem [{\citenamefont {Mansour~L}\ \emph {et~al.}(2021)\citenamefont {Mansour~L}, \citenamefont {Tian}, \citenamefont {Yeo}, \citenamefont {Cropley},\ and\ \citenamefont {Zalesky}}]{mansourl_high-resolution_2021}%
  \BibitemOpen
  \bibfield  {author} {\bibinfo {author} {\bibfnamefont {S.}~\bibnamefont {Mansour~L}}, \bibinfo {author} {\bibfnamefont {Y.}~\bibnamefont {Tian}}, \bibinfo {author} {\bibfnamefont {B.~T.}\ \bibnamefont {Yeo}}, \bibinfo {author} {\bibfnamefont {V.}~\bibnamefont {Cropley}},\ and\ \bibinfo {author} {\bibfnamefont {A.}~\bibnamefont {Zalesky}},\ }\href {https://doi.org/10.1016/j.neuroimage.2020.117695} {\bibfield  {journal} {\bibinfo  {journal} {NeuroImage}\ }\textbf {\bibinfo {volume} {229}},\ \bibinfo {pages} {117695} (\bibinfo {year} {2021})}\BibitemShut {NoStop}%
\bibitem [{\citenamefont {Mansour~L.}\ \emph {et~al.}(2023)\citenamefont {Mansour~L.}, \citenamefont {Di~Biase}, \citenamefont {Smith}, \citenamefont {Zalesky},\ and\ \citenamefont {Seguin}}]{mansour_l_connectomes_2023}%
  \BibitemOpen
  \bibfield  {author} {\bibinfo {author} {\bibfnamefont {S.}~\bibnamefont {Mansour~L.}}, \bibinfo {author} {\bibfnamefont {M.~A.}\ \bibnamefont {Di~Biase}}, \bibinfo {author} {\bibfnamefont {R.~E.}\ \bibnamefont {Smith}}, \bibinfo {author} {\bibfnamefont {A.}~\bibnamefont {Zalesky}},\ and\ \bibinfo {author} {\bibfnamefont {C.}~\bibnamefont {Seguin}},\ }\href {https://doi.org/10.1016/j.neuroimage.2023.120407} {\bibfield  {journal} {\bibinfo  {journal} {NeuroImage}\ }\textbf {\bibinfo {volume} {283}},\ \bibinfo {pages} {120407} (\bibinfo {year} {2023})}\BibitemShut {NoStop}%
\bibitem [{\citenamefont {Tournier}\ \emph {et~al.}(2019)\citenamefont {Tournier}, \citenamefont {Smith}, \citenamefont {Raffelt}, \citenamefont {Tabbara}, \citenamefont {Dhollander}, \citenamefont {Pietsch}, \citenamefont {Christiaens}, \citenamefont {Jeurissen}, \citenamefont {Yeh},\ and\ \citenamefont {Connelly}}]{tournier_mrtrix3_2019}%
  \BibitemOpen
  \bibfield  {author} {\bibinfo {author} {\bibfnamefont {J.-D.}\ \bibnamefont {Tournier}}, \bibinfo {author} {\bibfnamefont {R.}~\bibnamefont {Smith}}, \bibinfo {author} {\bibfnamefont {D.}~\bibnamefont {Raffelt}}, \bibinfo {author} {\bibfnamefont {R.}~\bibnamefont {Tabbara}}, \bibinfo {author} {\bibfnamefont {T.}~\bibnamefont {Dhollander}}, \bibinfo {author} {\bibfnamefont {M.}~\bibnamefont {Pietsch}}, \bibinfo {author} {\bibfnamefont {D.}~\bibnamefont {Christiaens}}, \bibinfo {author} {\bibfnamefont {B.}~\bibnamefont {Jeurissen}}, \bibinfo {author} {\bibfnamefont {C.-H.}\ \bibnamefont {Yeh}},\ and\ \bibinfo {author} {\bibfnamefont {A.}~\bibnamefont {Connelly}},\ }\href {https://doi.org/10.1016/j.neuroimage.2019.116137} {\bibfield  {journal} {\bibinfo  {journal} {NeuroImage}\ }\textbf {\bibinfo {volume} {202}},\ \bibinfo {pages} {116137} (\bibinfo {year} {2019})}\BibitemShut {NoStop}%
\bibitem [{\citenamefont {Jeurissen}\ \emph {et~al.}(2014)\citenamefont {Jeurissen}, \citenamefont {Tournier}, \citenamefont {Dhollander}, \citenamefont {Connelly},\ and\ \citenamefont {Sijbers}}]{jeurissen_multi-tissue_2014}%
  \BibitemOpen
  \bibfield  {author} {\bibinfo {author} {\bibfnamefont {B.}~\bibnamefont {Jeurissen}}, \bibinfo {author} {\bibfnamefont {J.-D.}\ \bibnamefont {Tournier}}, \bibinfo {author} {\bibfnamefont {T.}~\bibnamefont {Dhollander}}, \bibinfo {author} {\bibfnamefont {A.}~\bibnamefont {Connelly}},\ and\ \bibinfo {author} {\bibfnamefont {J.}~\bibnamefont {Sijbers}},\ }\href {https://doi.org/10.1016/j.neuroimage.2014.07.061} {\bibfield  {journal} {\bibinfo  {journal} {NeuroImage}\ }\textbf {\bibinfo {volume} {103}},\ \bibinfo {pages} {411} (\bibinfo {year} {2014})}\BibitemShut {NoStop}%
\bibitem [{\citenamefont {Dhollander}\ \emph {et~al.}(2016)\citenamefont {Dhollander}, \citenamefont {Raffelt},\ and\ \citenamefont {Connelly}}]{dhollander_unsupervised_2016}%
  \BibitemOpen
  \bibfield  {author} {\bibinfo {author} {\bibfnamefont {T.}~\bibnamefont {Dhollander}}, \bibinfo {author} {\bibfnamefont {D.}~\bibnamefont {Raffelt}},\ and\ \bibinfo {author} {\bibfnamefont {A.}~\bibnamefont {Connelly}},\ }in\ \href@noop {} {\emph {\bibinfo {booktitle} {{ISMRM} workshop on breaking the barriers of diffusion {MRI}}}},\ Vol.~\bibinfo {volume} {5}\ (\bibinfo  {publisher} {Lisbon},\ \bibinfo {year} {2016})\ \bibinfo {note} {number: 5}\BibitemShut {NoStop}%
\bibitem [{\citenamefont {Tournier}\ \emph {et~al.}(2010)\citenamefont {Tournier}, \citenamefont {Calamante}, \citenamefont {Connelly},\ and\ \citenamefont {{others}}}]{tournier_improved_2010}%
  \BibitemOpen
  \bibfield  {author} {\bibinfo {author} {\bibfnamefont {J.~D.}\ \bibnamefont {Tournier}}, \bibinfo {author} {\bibfnamefont {F.}~\bibnamefont {Calamante}}, \bibinfo {author} {\bibfnamefont {A.}~\bibnamefont {Connelly}},\ and\ \bibinfo {author} {\bibnamefont {{others}}},\ }in\ \href@noop {} {\emph {\bibinfo {booktitle} {Proceedings of the international society for magnetic resonance in medicine}}},\ Vol.\ \bibinfo {volume} {1670}\ (\bibinfo  {publisher} {Stockholm},\ \bibinfo {year} {2010})\BibitemShut {NoStop}%
\bibitem [{\citenamefont {Smith}\ \emph {et~al.}(2012)\citenamefont {Smith}, \citenamefont {Tournier}, \citenamefont {Calamante},\ and\ \citenamefont {Connelly}}]{smith_anatomically-constrained_2012}%
  \BibitemOpen
  \bibfield  {author} {\bibinfo {author} {\bibfnamefont {R.~E.}\ \bibnamefont {Smith}}, \bibinfo {author} {\bibfnamefont {J.-D.}\ \bibnamefont {Tournier}}, \bibinfo {author} {\bibfnamefont {F.}~\bibnamefont {Calamante}},\ and\ \bibinfo {author} {\bibfnamefont {A.}~\bibnamefont {Connelly}},\ }\href {https://doi.org/10.1016/j.neuroimage.2012.06.005} {\bibfield  {journal} {\bibinfo  {journal} {NeuroImage}\ }\textbf {\bibinfo {volume} {62}},\ \bibinfo {pages} {1924} (\bibinfo {year} {2012})}\BibitemShut {NoStop}%
\bibitem [{\citenamefont {Sanz~Leon}\ \emph {et~al.}(2013)\citenamefont {Sanz~Leon}, \citenamefont {Knock}, \citenamefont {Woodman}, \citenamefont {Domide}, \citenamefont {Mersmann}, \citenamefont {McIntosh},\ and\ \citenamefont {Jirsa}}]{sanz_leon_virtual_2013}%
  \BibitemOpen
  \bibfield  {author} {\bibinfo {author} {\bibfnamefont {P.}~\bibnamefont {Sanz~Leon}}, \bibinfo {author} {\bibfnamefont {S.~A.}\ \bibnamefont {Knock}}, \bibinfo {author} {\bibfnamefont {M.~M.}\ \bibnamefont {Woodman}}, \bibinfo {author} {\bibfnamefont {L.}~\bibnamefont {Domide}}, \bibinfo {author} {\bibfnamefont {J.}~\bibnamefont {Mersmann}}, \bibinfo {author} {\bibfnamefont {A.~R.}\ \bibnamefont {McIntosh}},\ and\ \bibinfo {author} {\bibfnamefont {V.}~\bibnamefont {Jirsa}},\ }\bibfield  {journal} {\bibinfo  {journal} {Frontiers in Neuroinformatics}\ }\textbf {\bibinfo {volume} {7}},\ \href {https://doi.org/10.3389/fninf.2013.00010} {10.3389/fninf.2013.00010} (\bibinfo {year} {2013})\BibitemShut {NoStop}%
\bibitem [{\citenamefont {Nunez}\ and\ \citenamefont {Srinivasan}(2006)}]{nunez_electric_2006}%
  \BibitemOpen
  \bibfield  {author} {\bibinfo {author} {\bibfnamefont {P.~L.}\ \bibnamefont {Nunez}}\ and\ \bibinfo {author} {\bibfnamefont {R.}~\bibnamefont {Srinivasan}},\ }\href {https://doi.org/10.1093/acprof:oso/9780195050387.001.0001} {\emph {\bibinfo {title} {Electric {Fields} of the {Brain}}}}\ (\bibinfo  {publisher} {Oxford University Press},\ \bibinfo {year} {2006})\BibitemShut {NoStop}%
\bibitem [{\citenamefont {Moosavi}\ \emph {et~al.}(2022)\citenamefont {Moosavi}, \citenamefont {Jirsa},\ and\ \citenamefont {Truccolo}}]{moosavi_critical_2022}%
  \BibitemOpen
  \bibfield  {author} {\bibinfo {author} {\bibfnamefont {S.~A.}\ \bibnamefont {Moosavi}}, \bibinfo {author} {\bibfnamefont {V.~K.}\ \bibnamefont {Jirsa}},\ and\ \bibinfo {author} {\bibfnamefont {W.}~\bibnamefont {Truccolo}},\ }\href {https://doi.org/10.1371/journal.pone.0272902} {\bibfield  {journal} {\bibinfo  {journal} {PLOS ONE}\ }\textbf {\bibinfo {volume} {17}},\ \bibinfo {pages} {e0272902} (\bibinfo {year} {2022})}\BibitemShut {NoStop}%
\bibitem [{\citenamefont {Houssaini}\ \emph {et~al.}(2020)\citenamefont {Houssaini}, \citenamefont {Bernard},\ and\ \citenamefont {Jirsa}}]{houssaini_epileptor_2020}%
  \BibitemOpen
  \bibfield  {author} {\bibinfo {author} {\bibfnamefont {K.~E.}\ \bibnamefont {Houssaini}}, \bibinfo {author} {\bibfnamefont {C.}~\bibnamefont {Bernard}},\ and\ \bibinfo {author} {\bibfnamefont {V.~K.}\ \bibnamefont {Jirsa}},\ }\bibfield  {journal} {\bibinfo  {journal} {eNeuro}\ }\textbf {\bibinfo {volume} {7}},\ \href {https://doi.org/10.1523/ENEURO.0485-18.2019} {10.1523/ENEURO.0485-18.2019} (\bibinfo {year} {2020}),\ \bibinfo {note} {section: Research Article: New Research}\BibitemShut {NoStop}%
\bibitem [{\citenamefont {Otsu}(1979)}]{otsu_threshold_1979}%
  \BibitemOpen
  \bibfield  {author} {\bibinfo {author} {\bibfnamefont {N.}~\bibnamefont {Otsu}},\ }\href@noop {} {\bibfield  {journal} {\bibinfo  {journal} {IEEE Transactions on Systems, Man, and Cybernetics}\ }\textbf {\bibinfo {volume} {9}},\ \bibinfo {pages} {23} (\bibinfo {year} {1979})}\BibitemShut {NoStop}%
\bibitem [{\citenamefont {Sun}\ \emph {et~al.}(2024)\citenamefont {Sun}, \citenamefont {Sohrabpour}, \citenamefont {Joseph}, \citenamefont {Worrell},\ and\ \citenamefont {He}}]{sun_seizure_2024}%
  \BibitemOpen
  \bibfield  {author} {\bibinfo {author} {\bibfnamefont {R.}~\bibnamefont {Sun}}, \bibinfo {author} {\bibfnamefont {A.}~\bibnamefont {Sohrabpour}}, \bibinfo {author} {\bibfnamefont {B.}~\bibnamefont {Joseph}}, \bibinfo {author} {\bibfnamefont {G.}~\bibnamefont {Worrell}},\ and\ \bibinfo {author} {\bibfnamefont {B.}~\bibnamefont {He}},\ }\href {https://doi.org/10.1002/advs.202405246} {\bibfield  {journal} {\bibinfo  {journal} {Advanced Science}\ }\textbf {\bibinfo {volume} {11}},\ \bibinfo {pages} {2405246} (\bibinfo {year} {2024})},\ \bibinfo {note} {\_eprint: https://onlinelibrary.wiley.com/doi/pdf/10.1002/advs.202405246}\BibitemShut {NoStop}%
\bibitem [{\citenamefont {Plummer}\ \emph {et~al.}(2019)\citenamefont {Plummer}, \citenamefont {Vogrin}, \citenamefont {Woods}, \citenamefont {Murphy}, \citenamefont {Cook},\ and\ \citenamefont {Liley}}]{plummer_interictal_2019}%
  \BibitemOpen
  \bibfield  {author} {\bibinfo {author} {\bibfnamefont {C.}~\bibnamefont {Plummer}}, \bibinfo {author} {\bibfnamefont {S.~J.}\ \bibnamefont {Vogrin}}, \bibinfo {author} {\bibfnamefont {W.~P.}\ \bibnamefont {Woods}}, \bibinfo {author} {\bibfnamefont {M.~A.}\ \bibnamefont {Murphy}}, \bibinfo {author} {\bibfnamefont {M.~J.}\ \bibnamefont {Cook}},\ and\ \bibinfo {author} {\bibfnamefont {D.~T.~J.}\ \bibnamefont {Liley}},\ }\href {https://doi.org/10.1093/brain/awz015} {\bibfield  {journal} {\bibinfo  {journal} {Brain}\ }\textbf {\bibinfo {volume} {142}},\ \bibinfo {pages} {932} (\bibinfo {year} {2019})}\BibitemShut {NoStop}%
\bibitem [{\citenamefont {Siu}(2026{\natexlab{a}})}]{siu_github_sim}%
  \BibitemOpen
  \bibfield  {author} {\bibinfo {author} {\bibfnamefont {P.~H.}\ \bibnamefont {Siu}},\ }\href {\url{https://github.com/Spokhim/sl_sim_framework}} {\bibinfo {title} {sl\_sim\_framework}} (\bibinfo {year} {2026}{\natexlab{a}}),\ \bibinfo {note} {gitHub repository, accessed 10 August 2026}\BibitemShut {NoStop}%
\bibitem [{\citenamefont {Siu}(2026{\natexlab{b}})}]{siu_github_sesl}%
  \BibitemOpen
  \bibfield  {author} {\bibinfo {author} {\bibfnamefont {P.~H.}\ \bibnamefont {Siu}},\ }\href {\url{https://github.com/Spokhim/sesl}} {\bibinfo {title} {sesl}} (\bibinfo {year} {2026}{\natexlab{b}}),\ \bibinfo {note} {gitHub repository, accessed 10 August 2026}\BibitemShut {NoStop}%
\end{thebibliography}%

\end{document}